\documentclass[10pt, conference, letterpaper]{IEEEtran}

\usepackage{tikz}
\usetikzlibrary{arrows.meta, positioning}

\usepackage[utf8]{inputenc}   
\usepackage[T1]{fontenc}    	
\usepackage{url}              		 
\usepackage{booktabs}       	
\usepackage{amsfonts}       	
\usepackage{nicefrac}       	  
\usepackage{microtype}      	
\usepackage[american]{babel}   	
\usepackage{comment}
\usepackage{svg}
\usepackage{graphicx}
\usepackage{color}
\usepackage{rotating}
\usepackage{tabularx}
\usepackage{pdflscape}

\usepackage{amsmath,amsthm,amssymb}
\usepackage{algorithm, algorithmic}
\usepackage{bbm,dsfont}
\usepackage{cancel}
\usepackage{enumerate, cases}
\usepackage{thmtools,thm-restate}
\usepackage{mathtools}

\usepackage[none]{hyphenat}
\usepackage{multirow}
\usepackage{makecell}
\usepackage{setspace}
\usepackage{pifont}
\usepackage{array}
\usepackage{enumitem}

\allowdisplaybreaks

\newcommand{\hide}[1] {}

\usepackage{silence}
\usepackage[capitalize]{cleveref}

\usepackage{graphicx}
\usepackage{subcaption}
\usepackage{float}   
\usepackage{stfloats}   

\begin{document}

	\title{PQC Migration: End-to-End Performance Profiling of Post Quantum Cryptographic Algorithms in mTLS}

    

	\author{
       \IEEEauthorblockN{Rahul, Manjesh K. Hanawal, Saravanan V}
        \IEEEauthorblockA{
        IIT Bombay, India}
        
        \and

        \IEEEauthorblockN{Sameer G. Kulkarni}
        \IEEEauthorblockA{
        IIT Gandhinagar, India}

        \and
   
        \IEEEauthorblockN{K Tirumaleswar Reddy}
        \IEEEauthorblockA{
        Nokia, India}
    }

	\maketitle
		\setlength{\columnsep}{0.2in}

	\begin{abstract}
      Emerging quantum computers threatens to break the classical key exchange algorithms. To safeguard against these threats, migration to new quantum-safe algorithms is underway and expected to be expedited. Though resistant to attacks, the quantum safe algorithms may impact the real-world performance due to their large key lengths. In this work, we analyze the end-to-end performance of the Transport Layer Security (TLS) protocol with the post-quantum (PQ) and hybrid cryptographic algorithms standardized by NIST. We develop a test setup to evaluate the performance of TLS connections between a client and a server under various network conditions. We focus on the impact of the new algorithms on the speed (throughput, connection time) and sustainability (energy, CPU cycles) of TLS 1.3. Our analysis demonstrates that the standardized post-quantum algorithms do not introduce any measurable performance degradations in TLS 1.3, despite their larger key lengths. 
      However, hybrid algorithms introduce noticeable performance degradation, which becomes increasingly pronounced under adverse network conditions.
      Our analysis suggests that migration to hybrid algorithms should be ``network condition aware'' as poor network conditions can severely degrade performance in TLS 1.3. We provide anonymised link to the code and a web interface to analyse performance of various combination of algorithms.
	\end{abstract}
	

	\IEEEpeerreviewmaketitle

	\section{Introduction}
	\label{sec:introduction}
	Internet security relies on various cryptographic algorithms to 
ensure the confidentiality, integrity, and authenticity of data. 
For decades, Rivest–Shamir–Adleman (RSA) and elliptic curve cryptography have been protecting everything from online banking to private messages. 
The security of these public-key cryptographic systems relies on the assumption that specific mathematical problems, namely integer factorization and the elliptic curve discrete logarithm problem, are hard to solve using conventional digital computers. However, quantum computers threaten this as they can solve hard problems much more efficiently.
Hence, there is an urgent need to migrate to cryptographic algorithms that remain secure even when the adversary uses a quantum computer. 

Key exchange algorithms are essential components for ensuring secure communication over the Internet. In August 2024, NIST released the first set of standards for post-quantum cryptography (PQC), which included Module-Lattice-based Key-Encapsulation-Mechanism, ML-KEM (FIPS 203) for key exchange  \cite{NIST:FIPS203}. Further, the Internet Engineering Task Force (IETF) is addressing the practical challenges of deploying post-quantum algorithms in TLS 1.3~\cite{rfc9958}, arising out of their larger public-key and ciphertext sizes. While post-quantum (PQ) key exchange algorithms can be used on their own, the IETF drafts \cite{IETF:TLS_MLKEM,IETF:UTA_PQC_APP} also propose using a \textit{hybrid} approach which combines traditional and PQ key exchange algorithms to generate the key. The hybrid algorithms are expected to mitigate the unknown vulnerabilities of PQ algorithms, while leveraging the time-tested strengths of traditional algorithms.

A natural question arises in PQC migrations: \textit{What will the transition to post-quantum key exchange algorithms actually cost us in terms of performance in the real world? And are they sustainable?} We are interested in the cost in terms of performance and energy consumption. For instance, when we replace a classical key exchange algorithm like X25519, which has a 32-byte key by a PQ algorithm like ML-KEM-1024, which has a 1,568-byte key, how will it impact connection speeds? Will real-world connections suffer from packet loss and latency? How are connection speeds affected for different values of these parameters? And perhaps more importantly, how does t affect overall energy consumption when all endpoints switch to PQ algorithms?

The IETF drafts recommend using ML-KEM-512, ML-KEM-768, and ML-KEM-1024 as post-quantum key exchange algorithms. Additionally, they suggest hybrid variants: X25519 combined with ML-KEM-768, SECP256r1 combined with ML-KEM-768, and SECP384r1 combined with ML-KEM-1024. In this paper, we benchmark these six algorithms against three widely used classical algorithms: X25519, SECP256r1, and SECP384r1.

We build a testing framework to evaluate the end-to-end performance of TLS connections. We use the cryptographic libraries provided by the Open Quantum Safe (OQS) \cite{OQSProvider} project in the OpenSSL implementation of TLS 1.3 \cite{OpenSSL}.
We measure throughput (the number of TLS handshakes per second), latency (time taken to establish a connection), CPU cycles, and Energy consumed. The first two metrics quantify speed while the other two quantify sustainability.

PQ and hybrid key exchange algorithms have larger public key lengths that can exceed the network's Maximum Transmission Unit (MTU), causing the TLS ClientHello message to fragment across multiple packets. This can degrade performance and increase energy consumption, with the impact amplified by link losses and network delays. Hence, we vary MTU size, link loss, and network delay, and measure their impact on the performance of TLS.

\begin{figure}[!htbp]
\centering

    \includegraphics[width=\columnwidth]{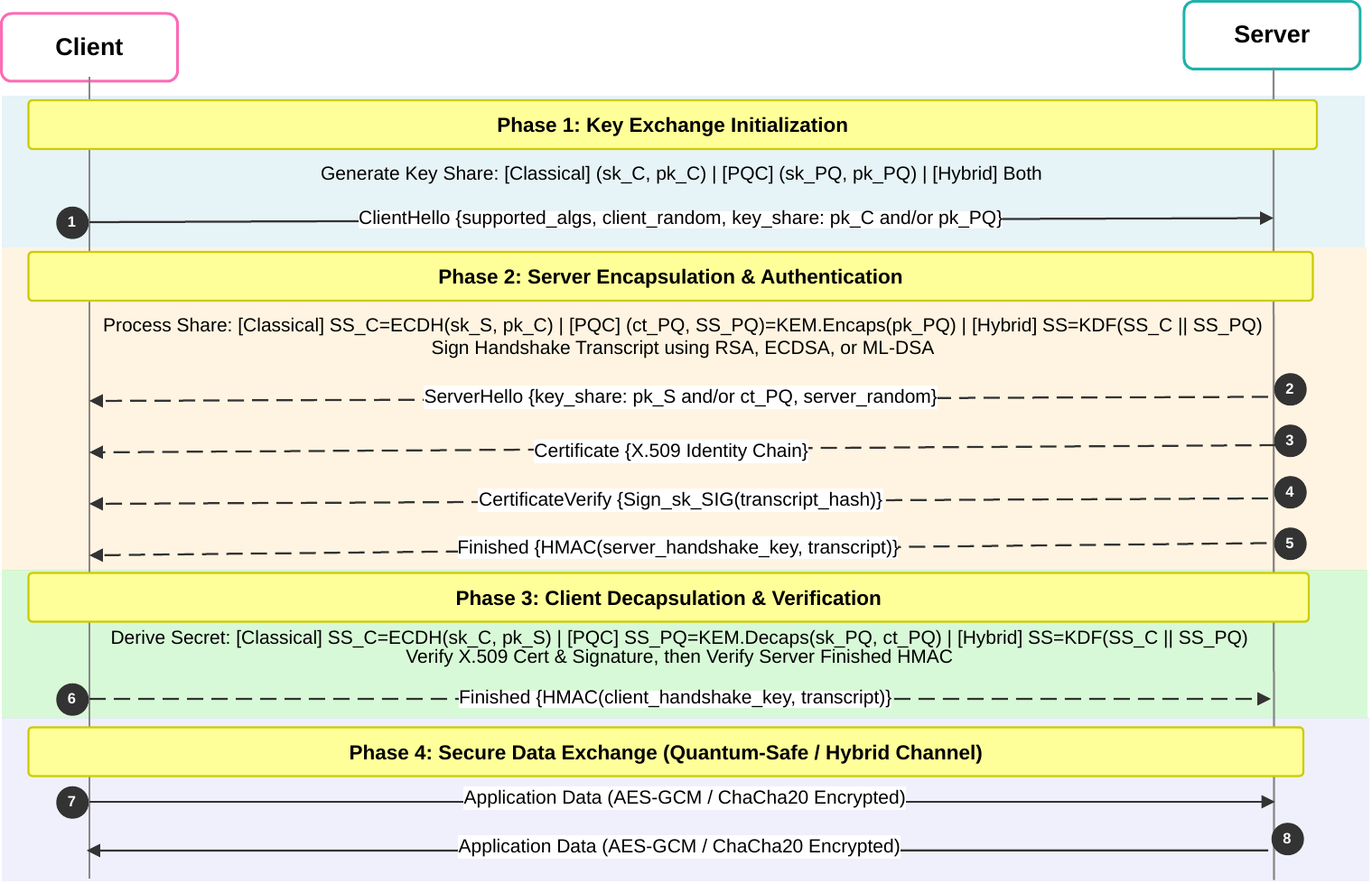}
    \caption{TLS~1.3 message flow }
    \label{fig:unified_tls_flow}
\end{figure}

Our analysis shows that substituting the classical algorithms with PQ algorithms in TLS 1.3 does not degrade performance in a ``typical'' network configuration. However, the use of hybrid algorithms results in noticeable performance degradation, with the impact becoming more pronounced as network conditions deteriorate (with high loss rates).
In summary, our contributions are as follows:
\begin{itemize}
    \item  We develop a unified framework to assess the performance and energy costs associated with migrating from classical cryptographic algorithms to the standardized post-quantum and hybrid alternatives within TLS 1.3.
    \item We evaluate the performance impact across a range of network conditions and quantify the computational, latency, and energy overheads introduced by PQ and hybrid cryptographic algorithms (\S~\ref{sec:Eval}).
    \item We developed a configurable TLS 1.3 evaluation platform that enables users to benchmark and compare the performance of arbitrary combinations of classical, PQ, and hybrid cryptographic algorithms.
    The interface is available at \cite{LiveBenchmark}.
    \end{itemize}

\noindent
\textbf{Organization}: Section~\ref{subsec:relatedwork} describes related work. In Section~\ref{sec:Algorithms}, we briefly discuss the key exchange algorithms whose performance was measured. In Section~\ref{sec:Setup}, we discuss our experimental setup, and discuss various metrics used for performance profiling in Section~\ref{sec:Metrics}. In Section~\ref{sec:experiments}, we discuss the observations and give conclusions in Section~\ref{sec:conclusion}. 



\subsection{Related work}
\label{subsec:relatedwork}
Several works have benchmarked the performance of PQ algorithms. Google \cite{Google:PQCExperiment:2016} and Cloudflare \cite{Cloudflare:PQC:2022} provided early performance analysis of TLS with hybrid algorithms by performing experiments over the Internet. Our work is closer to \cite{Paquin:BenchmarkingPQTLS:2020} and \cite{Sosnowski:PQCTLSPerformance:2023}, which develop a framework for running TLS with PQ algorithms by emulating network conditions. However, they predate NIST's announcement of standardized algorithms and compare various candidate algorithms. The recent study \cite{Nagy:MLKEMPerformance:2023} focuses on NIST-standardized algorithms but reports only the number of connections per second and overhead size under ideal scenarios. 
Several works have also considered the energy performance of PQ algorithms \cite{Schoffel:EnergyIoT:2021}, \cite{Roma:EnergyEfficiency:2021}, \cite{Barton:PQC_TLS_Constrained:2022},
\cite{Tasopoulos:EnergyEval:2023}. These works predate NIST's announcement of standardized algorithms and focus on comparing the energy performance of lattice-based, code-based, and isogeny-based families of PQ algorithms. The recent work \cite{Patterson:EnergyFramework:2025}
focuses on the energy usage of ML-KEM on Raspberry Pi devices. 

Our work differs from the above works in the focus on both speed and energy metrics of running TLS 1.3 with standardized PQ algorithms (both pure PQ and hybrid) when network conditions/parameters like network delay, packet loss and MTU size are varied. Our work reveals the impact of larger public keys in PQ and hybrid key exchange algorithms on TLS 1.3 performance. Following NIST standardization, several works have evaluated the finalized PQ key exchange algorithms in TLS across different scenarios. The work reported in \cite{Rios:QuantumSafeWeb}, \cite{Abbasi:PQCHeterogeneous:2025}, and \cite{GomezCambronero:LayeredTLS:2026} benchmarks PQ TLS throughput, latency, and MTU sensitivity in server-class environments. Works \cite{Kampanakis:DataHeavyTLS:2024} and \cite{Chou:PQCertTTFB:2026} study the effect of large PQ handshake and certificate sizes on connection completion time. For constrained and mobile deployments, \cite{Fitzgibbon:ConstrainedPQC:2024} and \cite{Claeys:TLS6TiSCH:2021} examine PQC on resource-limited hardware and fragmented networks. In contrast, \cite{Hoque:5GPQC:2025}, \cite{Setyowati:MobileIIoT:2025}, and \cite{Balaji:FinancialPQC:2026} study PQC in 5G, IIoT, and financial infrastructure settings. None of these works quantifies the retransmission overhead that TCP-level fragmentation introduces across NIST-standardized algorithms in the presence of packet loss, nor do they provide deployment guidance grounded in measured fragmentation thresholds. In this work, we specifically address these gaps.













    \section{Background and Relevant Algorithms}
    \label{sec:Algorithms}
\hide{
We assess the implications of transitioning from classical Elliptic-Curve Diffie--Hellman (ECDH) to post-quantum key encapsulation mechanisms (ML-KEM) for three cryptographic paradigms: classical, post-quantum, and hybrid under varied network conditions. 
}
We consider three cryptographic modes of operation in TLS 1.3, namely, classical, post-quantum, and hybrid.
Fig.~\ref{fig:unified_tls_flow} illustrates the TLS~1.3 handshake~\cite{RFC8446}, which is structurally identical across all three paradigms. First, the client advertises its key share in \texttt{ClientHello}, followed by the server's response in \texttt{ServerHello} and its certificate, and both sides derive a shared secret before exchanging \texttt{Finished} messages. What differs across configurations is solely the key exchange mechanism embedded within this flow. To evaluate the impact of both key exchange and authentication protocol migration, we evaluate two authentication configurations: \texttt{RSA-PSS (rsa\_pss\_rsae\_sha256)}~\cite{RFC8017}, reflecting current deployment practice, and \texttt{ML-DSA-44}~\cite{NIST:FIPS204}, a NIST-standardized post-quantum signature scheme. This dual-authentication design allows us to isolate the effect of key exchange from that of the signature algorithm.
\subsection{Classical Algorithms}

Classical elliptic-curve Diffie-Hellman (ECDH) key exchange algorithms serve as the performance baseline, representing current operational TLS deployments. X25519~\cite{RFC7748} uses a compact 32-byte public key and results in a total handshake size of 4,345 bytes.
Its small footprint allows the key exchange to fit within a single TCP packet, making it well-suited for latency-sensitive environments. The SECP256r1 (P-256) curve~\cite{NIST:SP800-186} increases the public key size to 65 bytes and produces a handshake of 4,411 bytes.
The SECP384r1 (P-384) configuration~\cite{NIST:SP800-186} further increases the public key size to 97 bytes and the overall handshake size to 4,475 bytes. Its higher security level incurs a computational cost approximately 2--2.5$\times$ that of SECP256r1 due to operations over a larger finite field.

\subsection{Post-Quantum Algorithms}

PQ configurations rely exclusively on lattice-based ML-KEM schemes~\cite{NIST:FIPS203}, providing quantum-resistant key exchange.
ML-KEM-512 uses an 800-byte public key and a 768-byte ciphertext, increasing the ClientHello size to 1,050 bytes. Despite this increase, efficient polynomial arithmetic offers a computational advantage, making it a lightweight PQ option. ML-KEM-768 increases the public key and ciphertext sizes to 1,184 bytes and 1,088 bytes, respectively, with a measured ClientHello of 1,434 bytes. 
ML-KEM-1024 further increases both public key and ciphertext sizes to 1,568 bytes each, yielding the largest pure PQ ClientHello of 1,818 bytes. 
Table~\ref{tab:alg_params} has the key sizes, ciphertext sizes, and total TLS~1.3 handshake sizes for all evaluated configurations.

\begin{table}[t]
  \centering
  \caption{Algorithm parameters and measured TLS~1.3 ClientHello (CH), ServerHello (SH), and total handshake sizes.}
  \label{tab:alg_params}
  \resizebox{\columnwidth}{!}{%
  \begin{tabular}{|l||c|c||c|c||c|}
    \hline
    \textbf{Algorithm} & \textbf{Key Size (B)} & \textbf{Ciphertext (B)} & \textbf{CH (B)} & \textbf{SH (B)} & \textbf{Handshake (B)} \\
    \hline\hline
    X25519            & 32    & ---   & 288   & 127   & 4,345  \\ \hline
    SECP256r1             & 65    & ---   & 321   & 160   & 4,411  \\ \hline
    SECP384r1             & 97    & ---   & 353   & 192   & 4,475  \\
    \hline\hline
    ML-KEM-512        & 800   & 768   & 1,050 & 863   & 5,854  \\ \hline
    ML-KEM-768        & 1,184 & 1,088 & 1,434 & 1,183 & 6,614  \\ \hline
    ML-KEM-1024       & 1,568 & 1,568 & 1,818 & 1,663 & 7,414  \\
    \hline\hline
    X25519+ML-KEM-768   & 1,216 & 1,120 & 1,466 & 1,215 & 6,646  \\ \hline
    SECP256r1+ML-KEM-768    & 1,249 & 1,153 & 1,499 & 1,248 & 6,712  \\ \hline
    SECP384r1+ML-KEM-1024   & 1,665 & 1,665 & 1,915 & 1,760 & 7,544  \\
    \hline
  \end{tabular}%
  }
\end{table}

\subsection{Hybrid Algorithms}

Hybrid configurations combine the classical ECDH key exchange with ML-KEM~\cite{NIST:FIPS203,Stebila:HybridTLS}, deriving a shared secret as
\begin{equation*}
K = \text{HKDF}(K_{\text{ECDH}} \parallel K_{\text{ML-KEM}})
\end{equation*}
where $K_{\text{ECDH}}$ is the ECDH shared key, $K_{\text{ML-KEM}}$ is the ML-KEM shared key, and HKDF is a HMAC-based key derivation function~\cite{RFC8446}.
This approach ensures backward compatibility while providing quantum resistance, at the cost of increased handshake size and computational overhead.
The X25519 + ML-KEM-768 configuration~\cite{NIST:FIPS203,RFC7748,IETF:ECDHE_MLKEM} combines a 32-byte classical key share with the 1,184-byte ML-KEM-768 public key and 1,088-byte ciphertext, yielding a combined key size of 1,216 bytes, ciphertext of 1,120 bytes, and a total handshake of 6,646 bytes.
Replacing X25519 with P-256~\cite{NIST:FIPS203,NIST:SP800-186,Stebila:HybridTLS} increases the combined public key to 1,249 bytes and ciphertext to 1,153 bytes, resulting in a total handshake of 6,712 bytes, with slightly higher computational cost due to prime-field arithmetic. 
The SECP384r1 + ML-KEM-1024 configuration~\cite{NIST:FIPS203,NIST:SP800-186,Stebila:HybridTLS} represents the most resource-intensive hybrid option, with a combined public key of 1,665 bytes, ciphertext of 1,665 bytes, and the largest hybrid handshake of 7,544 bytes, spanning up to 7--8 TCP packets.

    \section{Evaluation}
    \label{sec:Eval}
Though the PQ algorithms have larger handshake sizes, their simpler mathematical operations (matrix multiplications vs. exponentiation) can alleviate some of the performance degradation. However, several other factors come into play, like Maximum Transmission Unit (MTU) size, network delay, and packet loss. Due to larger public key and ciphertext lengths, the key share may not fit within a single MTU and will need to be sent over multiple fragments, resulting in transmission of additional TCP segments. The successful delivery of these segments is also affected by packet loss and network delay on the link. We thus configure these parameters in the performance evaluation of TLS 1.3 with classical, PQ and hybrid algorithms. 

	\subsection{Experimental Setup}
    \label{sec:Setup}
	\label{sec:normal_dist}
    We build containerized client and server instances to establish the TLS connections. Both run Ubuntu~22.04 with
OpenSSL~3.5.0, liboqs~3.5.0, and OQS~Provider~0.7.0 are interconnected over a
dedicated Docker bridge network. 
TLS~1.3 is 
enforced on both endpoints via the \texttt{-tls1\_3} flag, 
and no downgrade is permitted. The KEM and signature algorithm are selected at runtime via environment variables, enabling dynamic configuration and reuse of the containers without rebuilds across experiments.

\noindent\textbf{Mutual TLS and Certificate Infrastructure}:
All handshakes use mutual TLS (mTLS), in which both the client and server validate each other's certificates on every new connection. A fresh root CA certificate (RSA-2048 or ML-DSA-44, matching the authentication configuration under evaluation) and corresponding server and client leaf certificates are generated for each run, ensuring no certificate state persists across trials. Certificates are shared between containers via a dedicated volume mounted on the host device. This design mirrors deployments in security-sensitive networks, such as
5G core network and enterprise infrastructure, where both endpoints must authenticate each other to ensure trust.

\noindent\textbf{TLS Session Configuration}:
The server runs \texttt{openssl s\_server} on port~443. 
The client container runs a fresh openssl \texttt{s\_client} process that issues the HTTP/1.0 GET request. This ensures that the measured handshake is a full TLS session with no session resumption.
The key exchange group is set per algorithm via the \texttt{-groups} flag; no cipher suite restrictions are applied beyond the TLS~1.3 default.

\noindent\textbf{Host Performance Isolation}:
To eliminate OS-level noise and scheduling artifacts, the host is configured before every campaign.
All CPU cores are locked to 3.4\,GHz by setting the governor to performance
mode. We also pin the minimum and maximum scaling frequencies to the same value. All CPU idle states beyond C0 are disabled to keep cores continuously active, and the Intel P-state performance ceiling is set to 100\%. Turbo Boost is disabled.
Network conditions are emulated using Linux \texttt{netem}, enabling reproducible latency and packet loss scenarios; this reduces throughput variance from approximately 6\% to below 2\%.

\noindent\textbf{Container Resource Controls}:
We also pin the client and server containers to run on dedicated CPU cores 1 and~2 respectively. 
Both containers are allocated
4\,GB memory with elevated scheduling: $2\times$ the default CPU share weight,
with real-time priority~99.



\begin{figure}[t]
  \centering
  \includegraphics[width=\columnwidth]{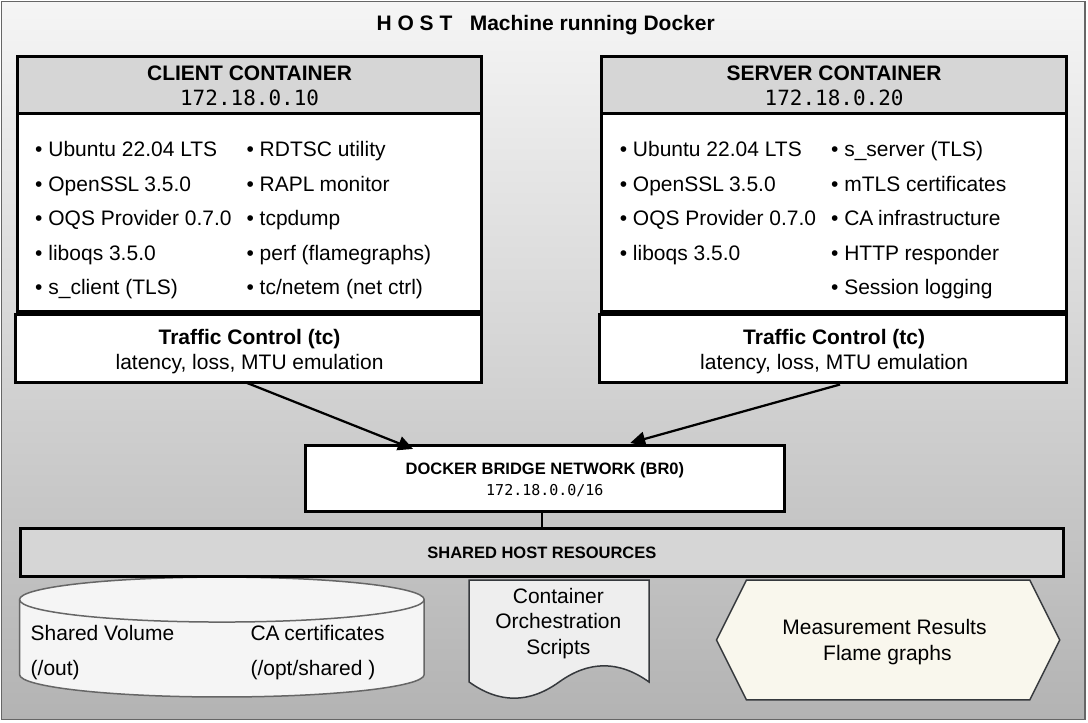} 
  \caption{Containerized testbed for PQC and hybrid TLS~1.3 benchmarking.}
  \label{fig:arch}
  \vspace{-12pt}
\end{figure}

	\subsection{Performance Metrics}
	\label{sec:Metrics}


To assess the practical viability of PQ TLS in production environments, we define four key performance metrics: throughput, connection time, energy per connection, and CPU cycles per connection. Each metric is evaluated over complete end-to-end TLS~1.3 handshakes with mutual authentication. This approach ensures that all reported results reflect realistic deployment conditions, incorporating both network effects and system-level interactions.
Below, we provide the details of all relevant metrics used in our evaluation:

\subsubsection{Throughput}
It is defined as the number of completed TLS connections per second. It is computed as the ratio of the total number of established connections and the elapsed measurement time. 

\subsubsection{Connection Time}
The connection time is defined as the total elapsed wall-clock time for a complete TLS connection, including the full handshake protocol exchange, certificate transmission, and the application-layer request-response cycle. Since each connection is a fresh single-handshake session with no resumption, the connection time equals the TLS handshake duration. Rather than relying on indirect system-level calculations (e.g., dividing total benchmark duration by throughput), latency is precisely measured at the hardware level using the \texttt{RDTSC} (Read Time-Stamp Counter) instruction, with the time recorded immediately before and after each call to the TLS connection function. This captures the total elapsed CPU cycles per connection, including both cryptographic computation and network round-trip times. The connection time (in seconds) is given by
\begin{equation}
T_{\text{conn}} = \frac{C_{\text{handshake}}}{f_{\text{CPU}}} 
\end{equation}
where $C_{\text{handshake}}$ is the number of elapsed CPU cycles per connection and $f_{\text{CPU}}$ is the processor timestamp frequency in Hz. Because network conditions are held constant across all experiments via \texttt{netem} and CPU frequency scaling is disabled, any systematic increase in $T_{\text{conn}}$ across algorithm configurations directly reflects the additional computational and transmission overhead imposed by the evaluated algorithms.

\subsubsection{Energy per Connection}
It is defined as the average energy consumed by the CPU package per TLS connection, measured in microjoules. It is obtained by reading the Intel RAPL \texttt{energy\_uj} counter for the CPU package domain (\texttt{intel-rapl:0}) via the Linux \texttt{sysfs} interface immediately before and after each connection. The difference gives the total package-level energy attributable to that connection, including both active computation and idle power consumed during network wait periods. Counter wraparound is handled explicitly. The per-connection value is averaged across all post-warm-up connections.

\subsubsection{CPU Cycles per Connection} It is defined as the CPU timestamp-counter ticks elapsed during a TLS connection, corresponding to the numerator $C_{\text{handshake}}$ in the connection time formula above. A serialized \texttt{CPUID}+\texttt{RDTSC} sequence is issued immediately before each connection, and a \texttt{RDTSCP}+\texttt{CPUID} sequence is issued immediately after, preventing out-of-order execution from contaminating either boundary. The cycle difference is averaged across all post-warm-up connections. Elapsed cycles include time spent waiting for network round-trips; because network conditions are fixed and identical across all algorithm configurations, systematic differences in this metric reflect PQC computational and transmission overhead.

\subsection{Measurement Methodology}

\subsubsection{Warm-up and Steady-State Execution}
Each experiment runs for a fixed duration of 60 seconds, divided into two phases. The first 15 connections constitute the \textit{warm-up phase} and are excluded from all measurements; this allows CPU caches to stabilize, kernel networking structures to initialize, and cryptographic execution paths to reach steady-state. All metrics reported correspond exclusively to \textit{post-warm-up} connections. Each scenario is repeated 50 times with complete system reinitialization between runs, ensuring statistical independence; results are reported as mean values with corresponding standard deviations.


\subsubsection{Connection Model}
Connections are established sequentially using OpenSSL's \texttt{s\_client} utility. For each connection, the client performs a full TLS 1.3 handshake with mutual authentication, sends a HTTP GET request, and terminates the connection once it receives the response from the server. 
\hide{
\begin{itemize}[noitemsep]
    \item Establishes a new TCP socket,
    \item Performs a full TLS~1.3 handshake with mutual authentication,
    \item Sends an HTTP GET request,
    \item Receives the server response, and
    \item Terminates the connection.
\end{itemize}
}
Each connection is independent, with no reuse of sessions or connection pooling. This design ensures that measurements capture true per-connection costs, including both cryptographic and network overhead.

All the metrics are collected concurrently during each experimental run. The measurement framework coordinates warm-up, data collection, and aggregation to produce consistent and reproducible results.



\noindent\textbf{Energy Analysis}:
Energy measurements obtained via Intel RAPL reflect total CPU package consumption, including both active computation and idle periods. Our analysis shows that CPU power consumption remains approximately constant—around 45~W with C-states disabled and 16~W with C-states enabled—independent of network latency.

As a result, total energy consumption scales linearly with connection duration:
\begin{equation}
E = P \times T
\end{equation}

Profiling reveals that nearly 97\% of CPU time is spent in idle/wait states, while only about 3\% is devoted to cryptographic computation. 
This demonstrates that network-induced delays dominate and CPU is used sporadically contributing to overall energy consumption. 
Due to hardware limitations, RAPL measurements cannot isolate per-process energy usage. Consequently, reported values include both active computation and idle power consumption during network wait periods.

    \subsection{Experiments}
	\label{sec:experiments}

\begin{figure*}[t!]
\centering

\begin{minipage}[c]{0.40\textwidth}
    \centering
    \textbf{(a)}\vspace{-12pt}\\
    \includegraphics[width=0.95\linewidth]{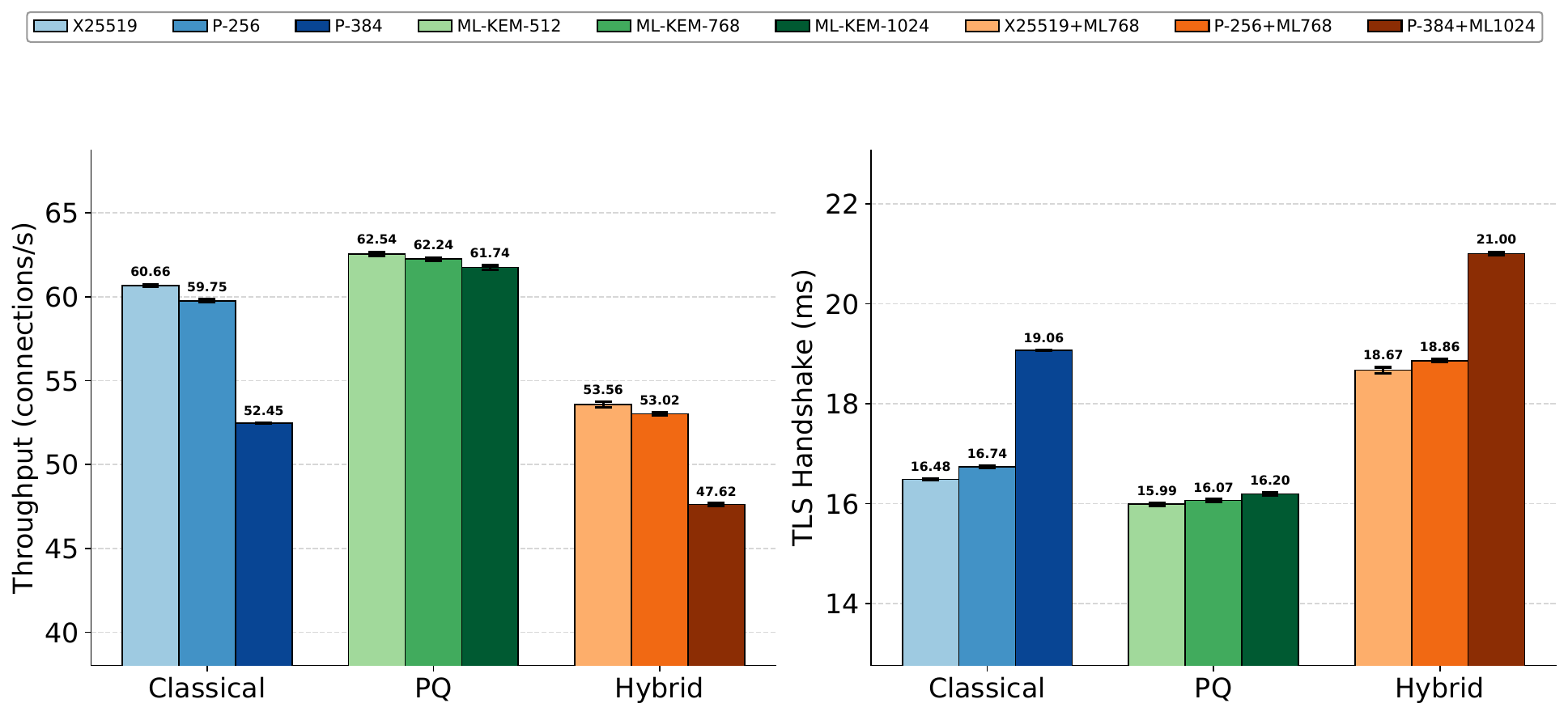}
\end{minipage}\hfill
\begin{minipage}[c]{0.55\textwidth}
    \centering
    \scriptsize
    \resizebox{\linewidth}{!}{%
    \begin{tabular}{|l|c|c|c|c|c|c|}
    \hline
    \multirow{2}{*}{Algorithm}
    & \multicolumn{2}{c|}{MTU=1200}
    & \multicolumn{2}{c|}{MTU=4500}
    & \multicolumn{2}{c|}{MTU=9000} \\
    \cline{2-7}
    & Thr & Conn (ms) & Thr & Conn (ms) & Thr & Conn (ms) \\
    \hline
    X25519               & 60.38 & 16.56 & 60.73 & 16.47 & 60.59 & 16.51 \\
    SECP256r1            & 59.78 & 16.73 & 59.95 & 16.68 & 60.08 & 16.65 \\
    SECP384r1            & 52.43 & 19.07 & 52.61 & 19.01 & 52.61 & 19.01 \\
    ML-KEM-512           & 61.90 & 16.16 & 62.07 & 16.11 & 62.16 & 16.09 \\
    ML-KEM-768           & 61.61 & 16.23 & 61.97 & 16.14 & 61.94 & 16.15 \\
    ML-KEM-1024          & 61.16 & 16.35 & 61.71 & 16.20 & 61.55 & 16.25 \\
    X25519-ML-KEM-768    & 53.90 & 18.55 & 53.75 & 18.61 & 53.69 & 18.63 \\
    SECP256r1-ML-KEM-768 & 53.10 & 18.83 & 53.22 & 18.79 & 52.79 & 18.95 \\
    SECP384r1-ML-KEM-1024 & 47.21 & 21.18 & 47.33 & 21.13 & 47.42 & 21.09 \\
    \hline
    \end{tabular}%
    }
\end{minipage}

\vspace{12pt}

\begin{minipage}[c]{0.40\textwidth}
    \centering
    \textbf{(b)}\vspace{-12pt}\\
    \includegraphics[width=0.95\linewidth]{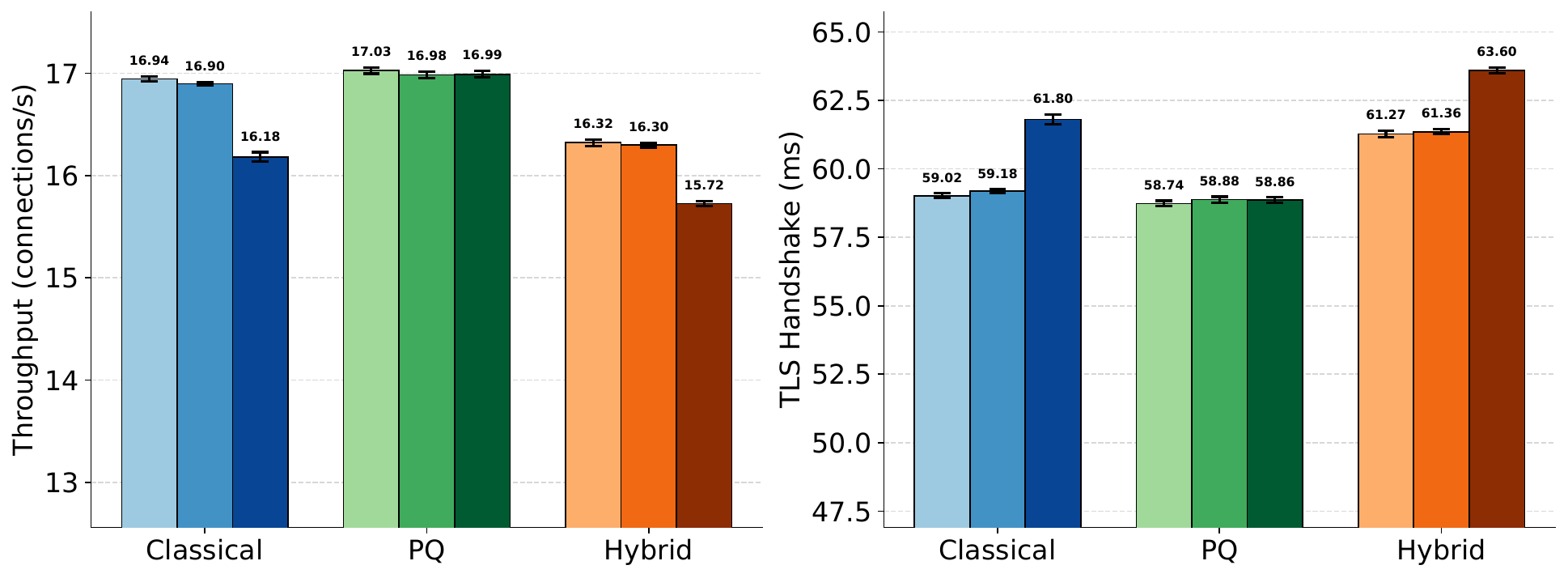}
\end{minipage}\hfill
\begin{minipage}[c]{0.55\textwidth}
    \centering
    \scriptsize
    \resizebox{\linewidth}{!}{%
    \begin{tabular}{|l|c|c|c|c|c|c|}
    \hline
    \multirow{2}{*}{Algorithm}
    & \multicolumn{2}{c|}{Delay=1ms}
    & \multicolumn{2}{c|}{Delay=10ms}
    & \multicolumn{2}{c|}{Delay=25ms} \\
    \cline{2-7}
    & Thr & Conn (ms) & Thr & Conn (ms) & Thr & Conn (ms) \\
    \hline
    X25519               & 40.03 & 24.98 & 9.73 & 97.76 & 4.25 & 216.79 \\
    SECP256r1            & 39.59 & 25.26 & 9.72 & 95.70 & 4.24 & 217.01 \\
    SECP384r1            & 35.23 & 28.38 & 9.45 & 98.06 & 4.19 & 221.35 \\
    ML-KEM-512           & 40.36 & 24.78 & 9.78 & 95.07 & 4.25 & 218.40 \\
    ML-KEM-768           & 40.25 & 24.85 & 9.77 & 94.76 & 4.25 & 216.50 \\
    ML-KEM-1024          & 40.22 & 24.86 & 9.75 & 95.15 & 4.24 & 216.84 \\
    X25519-ML-KEM-768    & 36.30 & 27.55 & 9.50 & 97.52 & 4.21 & 219.68 \\
    SECP256r1-ML-KEM-768 & 36.02 & 27.76 & 9.47 & 97.27 & 4.19 & 219.21 \\
    SECP384r1-ML-KEM-1024 & 34.71 & 28.81 & 9.24 & 99.84 & 4.15 & 222.06 \\
    \hline
    \end{tabular}%
    }
\end{minipage}


\begin{minipage}[c]{0.40\textwidth}
    \centering
    \textbf{(c)}\vspace{-12pt}\\
    \includegraphics[width=0.95\linewidth]{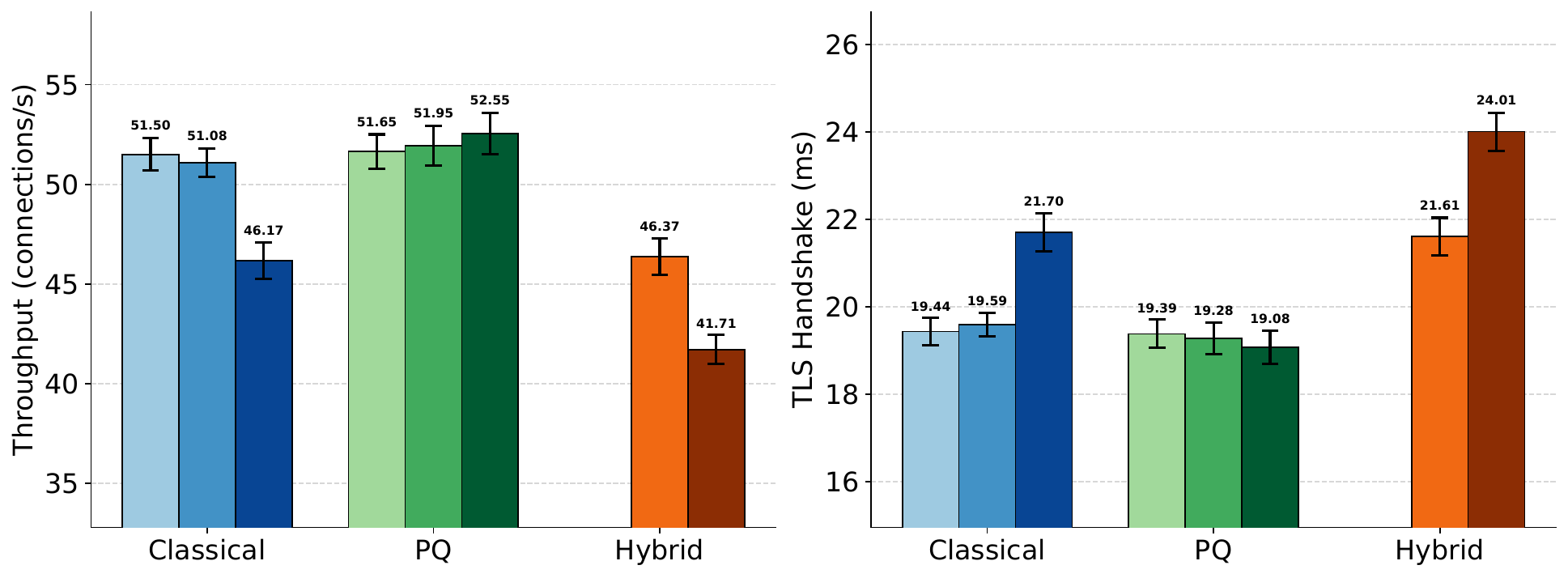}
\end{minipage}\hfill
\begin{minipage}[c]{0.55\textwidth}
    \centering
    \scriptsize
    \resizebox{\linewidth}{!}{%
    \begin{tabular}{|l|c|c|c|c|c|c|}
    \hline
    \multirow{2}{*}{Algorithm}
    & \multicolumn{2}{c|}{Loss=0.01\%}
    & \multicolumn{2}{c|}{Loss=1\%}
    & \multicolumn{2}{c|}{Loss=5\%} \\
    \cline{2-7}
    & Thr & Conn (ms) & Thr & Conn (ms) & Thr & Conn (ms) \\
    \hline
    X25519               & 59.78 & 12.37 & 22.87 & 39.44 & 5.27 & 201.71 \\
    SECP256r1            & 58.72 & 12.77 & 22.97 & 39.08 & 5.52 & 171.72 \\
    SECP384r1            & 51.78 & 15.06 & 21.44 & 42.36 & 5.09 & 196.32 \\
    ML-KEM-512           & 61.20 & 12.16 & 22.21 & 40.48 & 5.56 & 196.78 \\
    ML-KEM-768           & 60.82 & 12.20 & 23.28 & 38.60 & 5.52 & 176.87 \\
    ML-KEM-1024          & 60.20 & 12.38 & 21.37 & 42.76 & 4.47 & 222.35 \\
    X25519-ML-KEM-768    & 52.97 & 14.56 & 20.44 & 44.89 & 4.41 & 215.77 \\
    SECP256r1-ML-KEM-768 & 52.36 & 14.85 & 20.23 & 45.64 & 4.48 & 212.38 \\
    SECP384r1-ML-KEM-1024 & 46.72 & 16.93 & 20.56 & 44.47 & 4.86 & 218.07 \\
    \hline
    \end{tabular}%
    }
\end{minipage}

\caption{Throughput (conn/s) and connection time with RSA-2048 authentication. In each row, the bar chart (left) shows one representative operating point while the table (right) covers the full parameter sweep. (a) MTU sweep: bar chart at baseline (MTU=1500, zero delay, zero loss); table at MTU=1200, 4500, and 9000\,B. (b) Delay sweep: bar chart at 5\,ms one-way delay; table at 1, 10, and 25\,ms. (c) Loss sweep: bar chart at 0.1\% loss; table at 0.01\%, 1\%, and 5\%.}
\label{fig:thr-all}
\end{figure*}

\begin{figure*}[t!]
\centering

\begin{minipage}[c]{0.40\textwidth}
    \centering
    \textbf{(a)}\vspace{-12pt}\\
    \includegraphics[width=0.95\linewidth]{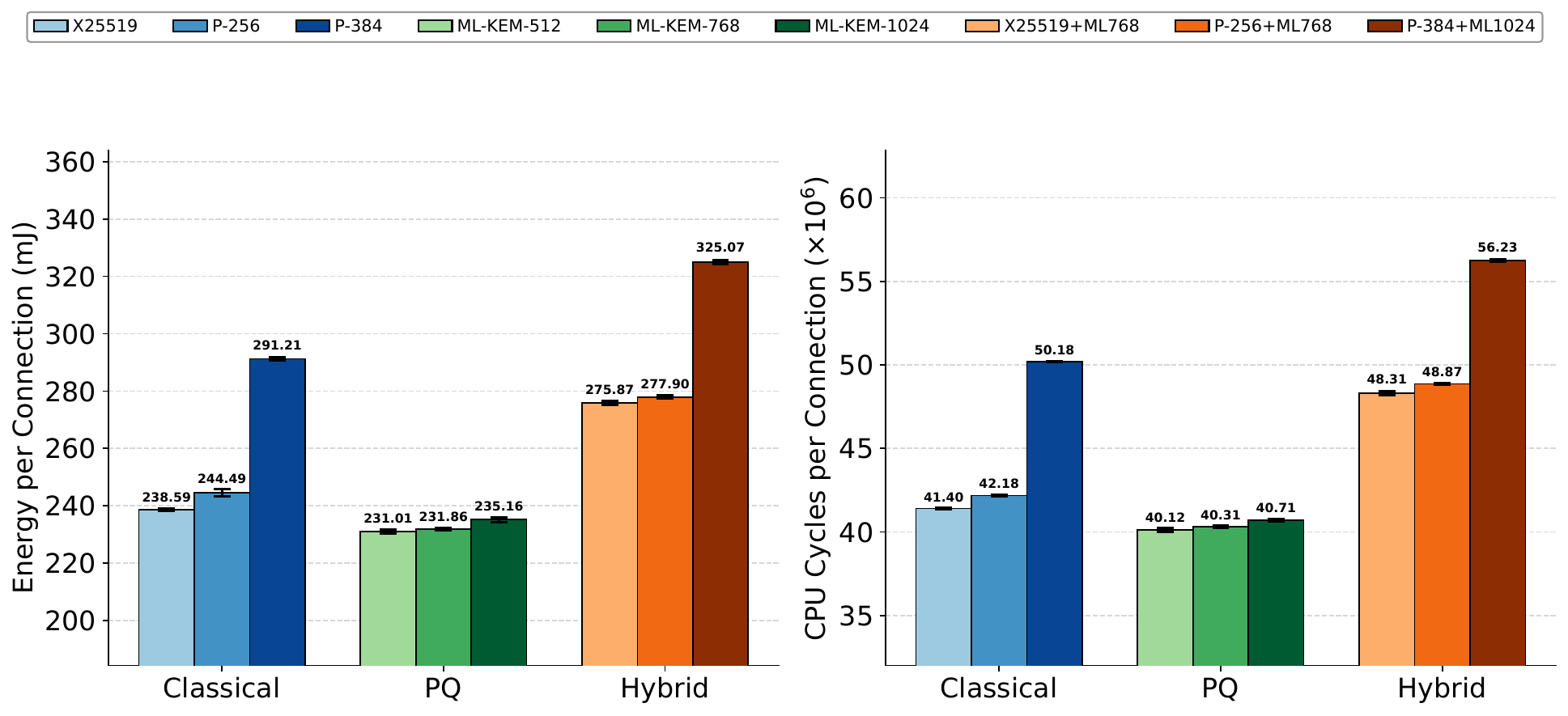}
\end{minipage}\hfill
\begin{minipage}[c]{0.55\textwidth}
    \centering
    \scriptsize
    \resizebox{\linewidth}{!}{%
    \begin{tabular}{|l|c|c|c|c|c|c|}
    \hline
    \multirow{2}{*}{Algorithm}
    & \multicolumn{2}{c|}{MTU=1200}
    & \multicolumn{2}{c|}{MTU=4500}
    & \multicolumn{2}{c|}{MTU=9000} \\
    \cline{2-7}
    & Energy (mJ) & CPU ($\times10^6$) & Energy (mJ) & CPU ($\times10^6$) & Energy (mJ) & CPU ($\times10^6$) \\
    \hline
    X25519               & 299.12 & 41.39 & 293.43 & 41.17 & 295.20 & 41.28 \\
    SECP256r1            & 302.27 & 41.99 & 297.28 & 41.90 & 297.29 & 41.81 \\
    SECP384r1            & 361.69 & 50.05 & 351.20 & 49.91 & 354.56 & 49.89 \\
    ML-KEM-512           & 291.30 & 40.37 & 284.82 & 40.27 & 286.23 & 40.16 \\
    ML-KEM-768           & 293.31 & 40.58 & 285.11 & 40.33 & 288.69 & 40.36 \\
    ML-KEM-1024          & 295.48 & 40.87 & 284.72 & 40.53 & 290.09 & 40.65 \\
    X25519-ML-KEM-768    & 337.43 & 48.22 & 342.14 & 48.45 & 341.04 & 48.40 \\
    SECP256r1-ML-KEM-768 & 346.19 & 48.99 & 343.43 & 48.85 & 362.65 & 49.29 \\
    SECP384r1-ML-KEM-1024 & 411.70 & 56.58 & 402.30 & 56.42 & 399.76 & 56.41 \\
    \hline
    \end{tabular}%
    }
\end{minipage}


\begin{minipage}[c]{0.40\textwidth}
    \centering
    \textbf{(b)}\vspace{-12pt}\\
    \includegraphics[width=0.95\linewidth]{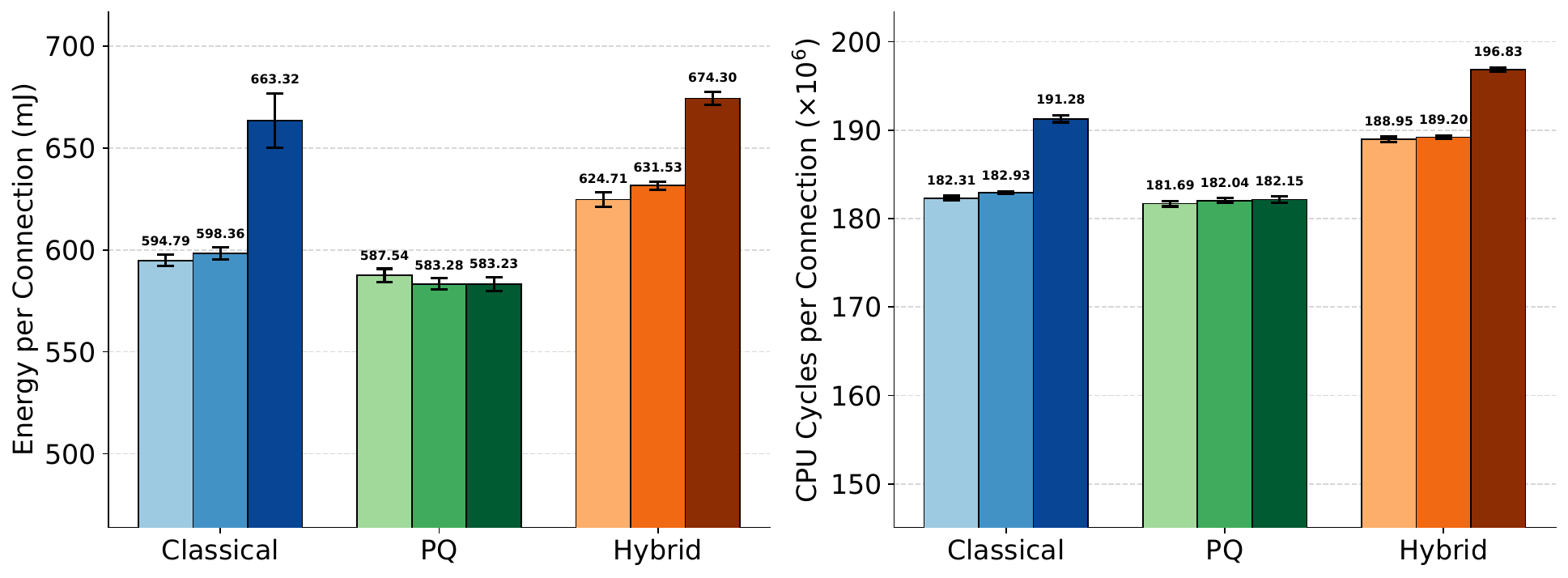}
\end{minipage}\hfill
\begin{minipage}[c]{0.55\textwidth}
    \centering
    \scriptsize
    \resizebox{\linewidth}{!}{%
    \begin{tabular}{|l|c|c|c|c|c|c|}
    \hline
    \multirow{2}{*}{Algorithm}
    & \multicolumn{2}{c|}{Delay=1ms}
    & \multicolumn{2}{c|}{Delay=10ms}
    & \multicolumn{2}{c|}{Delay=25ms} \\
    \cline{2-7}
    & Energy (mJ) & CPU ($\times10^6$) & Energy (mJ) & CPU ($\times10^6$) & Energy (mJ) & CPU ($\times10^6$) \\
    \hline
    X25519               & 432.64 & 69.40 & 1665.21 & 322.61 & 2523.83 & 736.39 \\
    SECP256r1            & 434.88 & 70.09 & 1735.26 & 323.13 & 2526.45 & 737.56 \\
    SECP384r1            & 493.36 & 80.26 & 1772.29 & 332.18 & 2582.73 & 745.71 \\
    ML-KEM-512           & 425.44 & 68.51 & 1515.11 & 321.43 & 2518.87 & 735.47 \\
    ML-KEM-768           & 428.05 & 68.72 & 1498.57 & 321.74 & 2492.92 & 736.27 \\
    ML-KEM-1024          & 427.85 & 68.92 & 1499.28 & 322.02 & 2504.79 & 735.76 \\
    X25519-ML-KEM-768    & 485.10 & 78.10 & 1528.66 & 329.88 & 2606.54 & 744.43 \\
    SECP256r1-ML-KEM-768 & 488.45 & 78.82 & 1435.42 & 330.61 & 2602.89 & 745.48 \\
    SECP384r1-ML-KEM-1024 & 529.86 & 82.54 & 1509.24 & 339.43 & 2641.79 & 754.06 \\
    \hline
    \end{tabular}%
    }
\end{minipage}


\begin{minipage}[c]{0.40\textwidth}
    \centering
    \textbf{(c)}\vspace{-12pt}\\
    \includegraphics[width=0.95\linewidth]{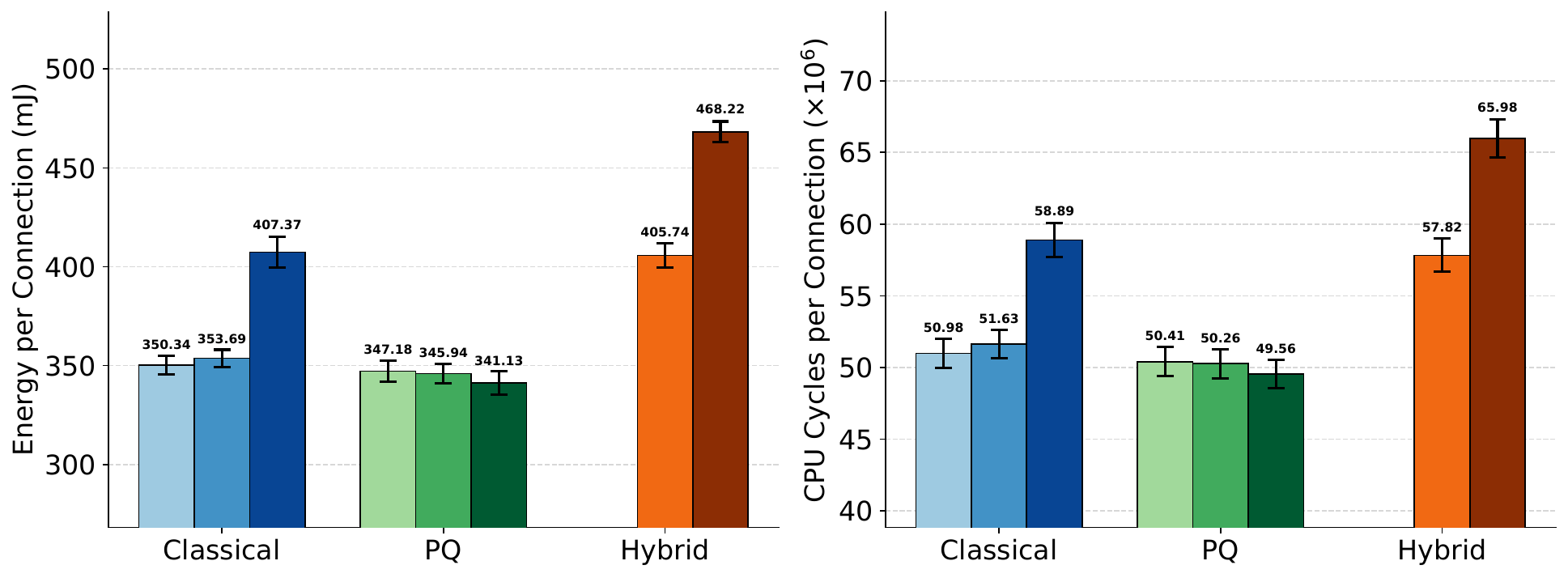}
\end{minipage}\hfill
\begin{minipage}[c]{0.55\textwidth}
    \centering
    \scriptsize
    \resizebox{\linewidth}{!}{%
    \begin{tabular}{|l|c|c|c|c|c|c|}
    \hline
    \multirow{2}{*}{Algorithm}
    & \multicolumn{2}{c|}{Loss=0.01\%}
    & \multicolumn{2}{c|}{Loss=1\%}
    & \multicolumn{2}{c|}{Loss=5\%} \\
    \cline{2-7}
    & Energy (mJ) & CPU ($\times10^6$) & Energy (mJ) & CPU ($\times10^6$) & Energy (mJ) & CPU ($\times10^6$) \\
    \hline
    X25519               & 293.62 & 42.07 & 575.82 & 133.91 & 1661.04 & 609.37 \\
    SECP256r1            & 296.89 & 43.15 & 572.08 & 132.86 & 1612.02 & 574.62 \\
    SECP384r1            & 363.16 & 50.97 & 636.29 & 143.99 & 1705.59 & 628.63 \\
    ML-KEM-512           & 287.07 & 41.05 & 568.74 & 137.19 & 1634.00 & 582.63 \\
    ML-KEM-768           & 287.87 & 41.37 & 552.14 & 130.82 & 1627.77 & 585.32 \\
    ML-KEM-1024          & 290.25 & 41.87 & 600.58 & 145.35 & 1764.53 & 724.33 \\
    X25519-ML-KEM-768    & 343.18 & 49.10 & 647.30 & 152.34 & 1820.77 & 725.63 \\
    SECP256r1-ML-KEM-768 & 346.68 & 49.84 & 650.72 & 153.84 & 1813.28 & 716.96 \\
    SECP384r1-ML-KEM-1024 & 404.89 & 57.53 & 671.83 & 149.92 & 1749.74 & 666.90 \\
    \hline
    \end{tabular}%
    }
\end{minipage}

\caption{Energy (mJ) and CPU cycles ($\times10^6$) per connection with RSA-2048 authentication. Layout as Fig.~\ref{fig:thr-all}: (a) MTU sweep (bar chart at MTU=1500 baseline; table at MTU=1200, 4500, 9000\,B), (b) delay sweep (bar chart at 5\,ms; table at 1, 10, 25\,ms), (c) loss sweep (bar chart at 0.1\%; table at 0.01\%, 1\%, 5\%).}
\label{fig:energy-all}
\end{figure*}


\begin{figure*}[t!]
\centering

\begin{minipage}[c]{0.40\textwidth}
    \centering
    \textbf{(a)}\vspace{-12pt}\\
    \includegraphics[width=0.95\linewidth]{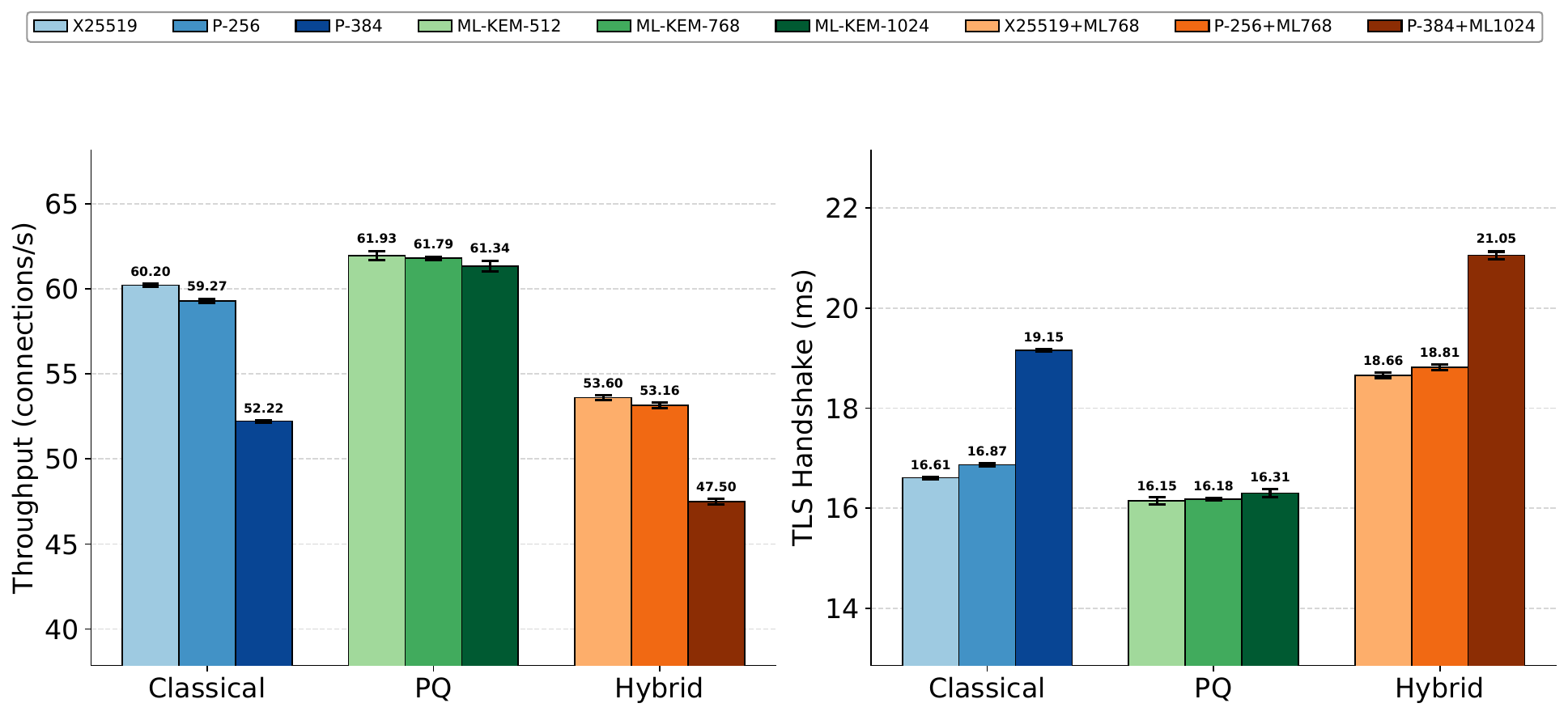}
\end{minipage}\hfill
\begin{minipage}[c]{0.55\textwidth}
    \centering
    \scriptsize
    \resizebox{\linewidth}{!}{%
    \begin{tabular}{|l|c|c|c|c|c|c|}
    \hline
    \multirow{2}{*}{Algorithm}
    & \multicolumn{2}{c|}{MTU=1200}
    & \multicolumn{2}{c|}{MTU=4500}
    & \multicolumn{2}{c|}{MTU=9000} \\
    \cline{2-7}
    & Thr & Conn (ms) & Thr & Conn (ms) & Thr & Conn (ms) \\
    \hline
    X25519               & 60.27 & 16.59 & 60.02 & 16.66 & 59.98 & 16.67 \\
    SECP256r1            & 59.30 & 16.86 & 59.48 & 16.81 & 59.21 & 16.89 \\
    SECP384r1            & 52.18 & 19.16 & 52.25 & 19.14 & 52.30 & 19.12 \\
    ML-KEM-512           & 61.69 & 16.21 & 61.28 & 16.32 & 61.01 & 16.39 \\
    ML-KEM-768           & 61.45 & 16.27 & 61.41 & 16.28 & 60.77 & 16.46 \\
    ML-KEM-1024          & 61.11 & 16.36 & 61.27 & 16.32 & 60.84 & 16.44 \\
    X25519-ML-KEM-768    & 53.73 & 18.61 & 53.89 & 18.56 & 53.73 & 18.61 \\
    SECP256r1-ML-KEM-768 & 53.09 & 18.84 & 53.27 & 18.77 & 53.31 & 18.76 \\
    SECP384r1-ML-KEM-1024 & 47.33 & 21.13 & 47.30 & 21.14 & 47.30 & 21.14 \\
    \hline
    \end{tabular}%
    }
\end{minipage}

\vspace{12pt}

\begin{minipage}[c]{0.40\textwidth}
    \centering
    \textbf{(b)}\vspace{-12pt}\\
    \includegraphics[width=0.95\linewidth]{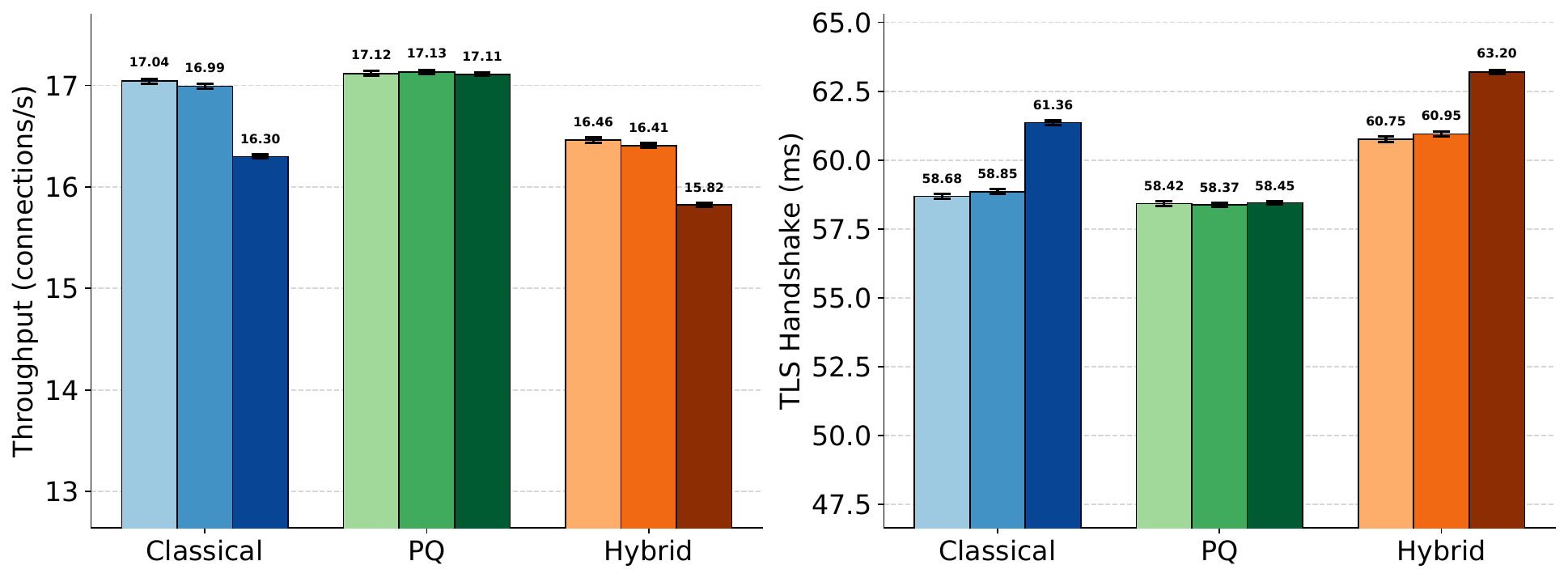}
\end{minipage}\hfill
\begin{minipage}[c]{0.55\textwidth}
    \centering
    \scriptsize
    \resizebox{\linewidth}{!}{%
    \begin{tabular}{|l|c|c|c|c|c|c|}
    \hline
    \multirow{2}{*}{Algorithm}
    & \multicolumn{2}{c|}{Delay=1ms}
    & \multicolumn{2}{c|}{Delay=10ms}
    & \multicolumn{2}{c|}{Delay=25ms} \\
    \cline{2-7}
    & Thr & Conn (ms) & Thr & Conn (ms) & Thr & Conn (ms) \\
    \hline
    X25519               & 40.61 & 24.63 & 9.79 & 94.31 & 4.23 & 218.86 \\
    SECP256r1            & 40.27 & 24.84 & 9.76 & 95.03 & 4.24 & 217.63 \\
    SECP384r1            & 36.04 & 27.75 & 9.52 & 97.23 & 4.19 & 219.75 \\
    ML-KEM-512           & 41.05 & 24.36 & 9.82 & 94.28 & 4.25 & 217.91 \\
    ML-KEM-768           & 41.06 & 24.36 & 9.82 & 94.26 & 4.25 & 216.39 \\
    ML-KEM-1024          & 40.87 & 24.47 & 9.81 & 94.63 & 4.26 & 215.74 \\
    X25519-ML-KEM-768    & 37.56 & 26.63 & 9.58 & 96.18 & 4.22 & 218.14 \\
    SECP256r1-ML-KEM-768 & 37.39 & 26.75 & 9.57 & 96.51 & 4.22 & 218.26 \\
    SECP384r1-ML-KEM-1024 & 34.96 & 28.61 & 9.28 & 99.37 & 4.15 & 221.82 \\
    \hline
    \end{tabular}%
    }
\end{minipage}

\vspace{12pt}

\begin{minipage}[c]{0.40\textwidth}
    \centering
    \textbf{(c)}\vspace{-12pt}\\
    \includegraphics[width=0.95\linewidth]{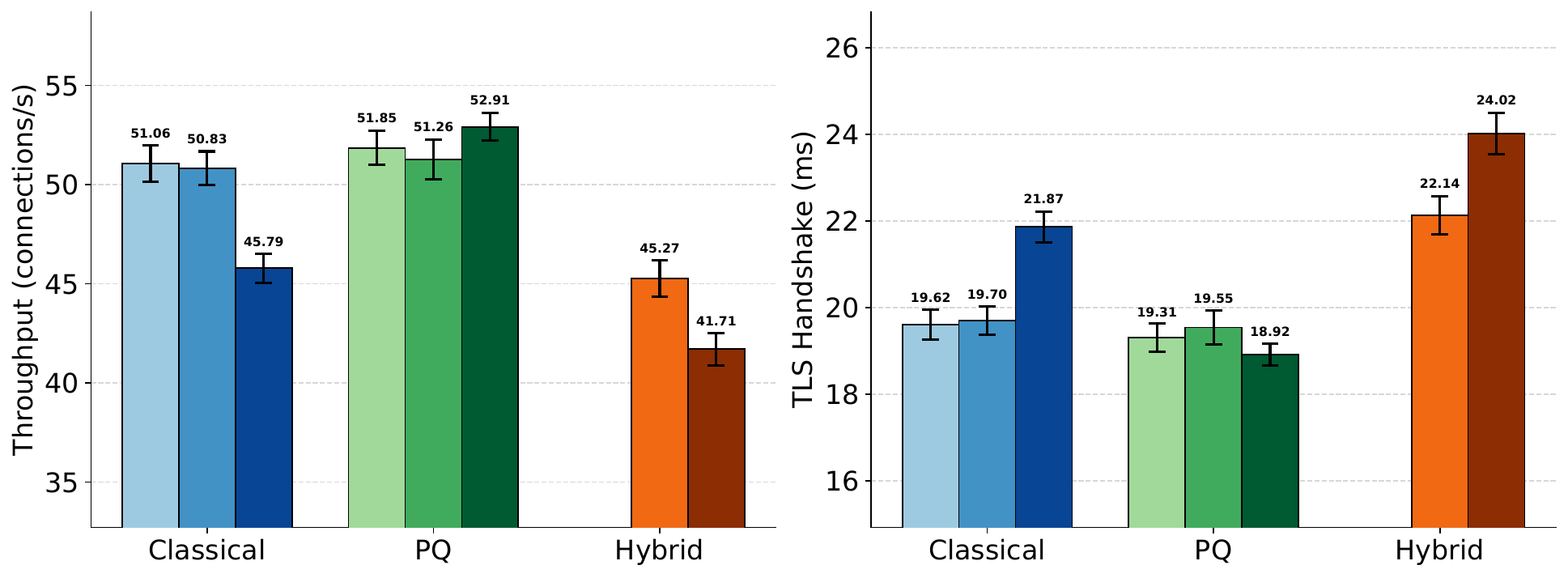}
\end{minipage}\hfill
\begin{minipage}[c]{0.55\textwidth}
    \centering
    \scriptsize
    \resizebox{\linewidth}{!}{%
    \begin{tabular}{|l|c|c|c|c|c|c|}
    \hline
    \multirow{2}{*}{Algorithm}
    & \multicolumn{2}{c|}{Loss=0.01\%}
    & \multicolumn{2}{c|}{Loss=1\%}
    & \multicolumn{2}{c|}{Loss=5\%} \\
    \cline{2-7}
    & Thr & Conn (ms) & Thr & Conn (ms) & Thr & Conn (ms) \\
    \hline
    X25519               & 59.09 & 12.64 & 22.29 & 40.72 & 5.49 & 178.73 \\
    SECP256r1            & 58.50 & 12.68 & 22.50 & 41.05 & 5.34 & 178.69 \\
    SECP384r1            & 51.52 & 15.23 & 20.91 & 43.65 & 5.35 & 183.51 \\
    ML-KEM-512           & 60.82 & 12.15 & 23.59 & 38.58 & 5.41 & 180.41 \\
    ML-KEM-768           & 60.55 & 12.39 & 22.13 & 41.04 & 5.23 & 181.65 \\
    ML-KEM-1024          & 60.01 & 12.34 & 22.70 & 40.19 & 4.59 & 211.72 \\
    X25519-ML-KEM-768    & 52.91 & 14.60 & 21.08 & 42.96 & 4.58 & 225.40 \\
    SECP256r1-ML-KEM-768 & 52.45 & 14.80 & 22.44 & 39.91 & 4.69 & 219.99 \\
    SECP384r1-ML-KEM-1024 & 46.83 & 16.85 & 19.46 & 47.90 & 4.47 & 230.94 \\
    \hline
    \end{tabular}%
    }
\end{minipage}

\caption{Throughput (conn/s) and connection time with ML-DSA-44 authentication. Layout as Fig.~\ref{fig:thr-all}: (a) MTU sweep, (b) delay sweep at 5\,ms, (c) loss sweep at 0.1\%.}
\label{fig:thr-all-mldsa44}
\end{figure*}


\begin{figure*}[t!]
\centering

\begin{minipage}[c]{0.40\textwidth}
    \centering
    \textbf{(a)}\vspace{-12pt}\\
    \includegraphics[width=0.95\linewidth]{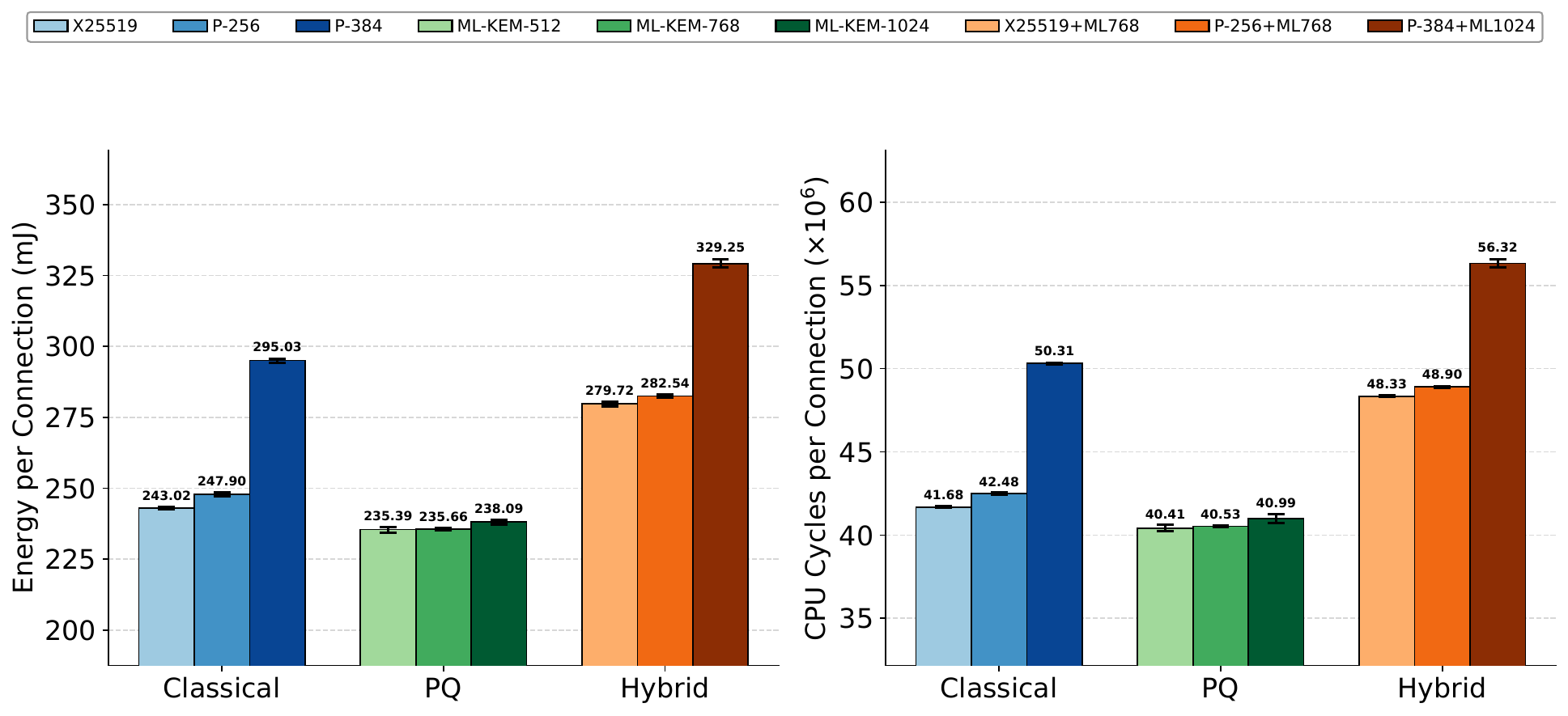}
\end{minipage}\hfill
\begin{minipage}[c]{0.55\textwidth}
    \centering
    \scriptsize
    \resizebox{\linewidth}{!}{%
    \begin{tabular}{|l|c|c|c|c|c|c|}
    \hline
    \multirow{2}{*}{Algorithm}
    & \multicolumn{2}{c|}{MTU=1200}
    & \multicolumn{2}{c|}{MTU=4500}
    & \multicolumn{2}{c|}{MTU=9000} \\
    \cline{2-7}
    & Energy (mJ) & CPU ($\times10^6$) & Energy (mJ) & CPU ($\times10^6$) & Energy (mJ) & CPU ($\times10^6$) \\
    \hline
    X25519               & 294.60 & 41.49 & 298.31 & 41.62 & 297.84 & 41.68 \\
    SECP256r1            & 299.77 & 42.28 & 300.78 & 42.21 & 302.88 & 42.37 \\
    SECP384r1            & 355.30 & 50.17 & 358.13 & 50.13 & 357.10 & 50.14 \\
    ML-KEM-512           & 287.49 & 40.40 & 292.38 & 40.58 & 299.44 & 40.79 \\
    ML-KEM-768           & 288.91 & 40.62 & 291.98 & 40.62 & 302.43 & 41.01 \\
    ML-KEM-1024          & 291.49 & 40.92 & 292.41 & 40.81 & 297.43 & 41.08 \\
    X25519-ML-KEM-768    & 343.91 & 48.31 & 347.25 & 48.37 & 342.40 & 48.47 \\
    SECP256r1-ML-KEM-768 & 347.54 & 48.95 & 349.42 & 48.85 & 345.68 & 48.95 \\
    SECP384r1-ML-KEM-1024 & 408.77 & 56.44 & 407.78 & 56.42 & 412.97 & 56.58 \\
    \hline
    \end{tabular}%
    }
\end{minipage}

\vspace{12pt}

\begin{minipage}[c]{0.40\textwidth}
    \centering
    \textbf{(b)}\vspace{-12pt}\\
    \includegraphics[width=0.95\linewidth]{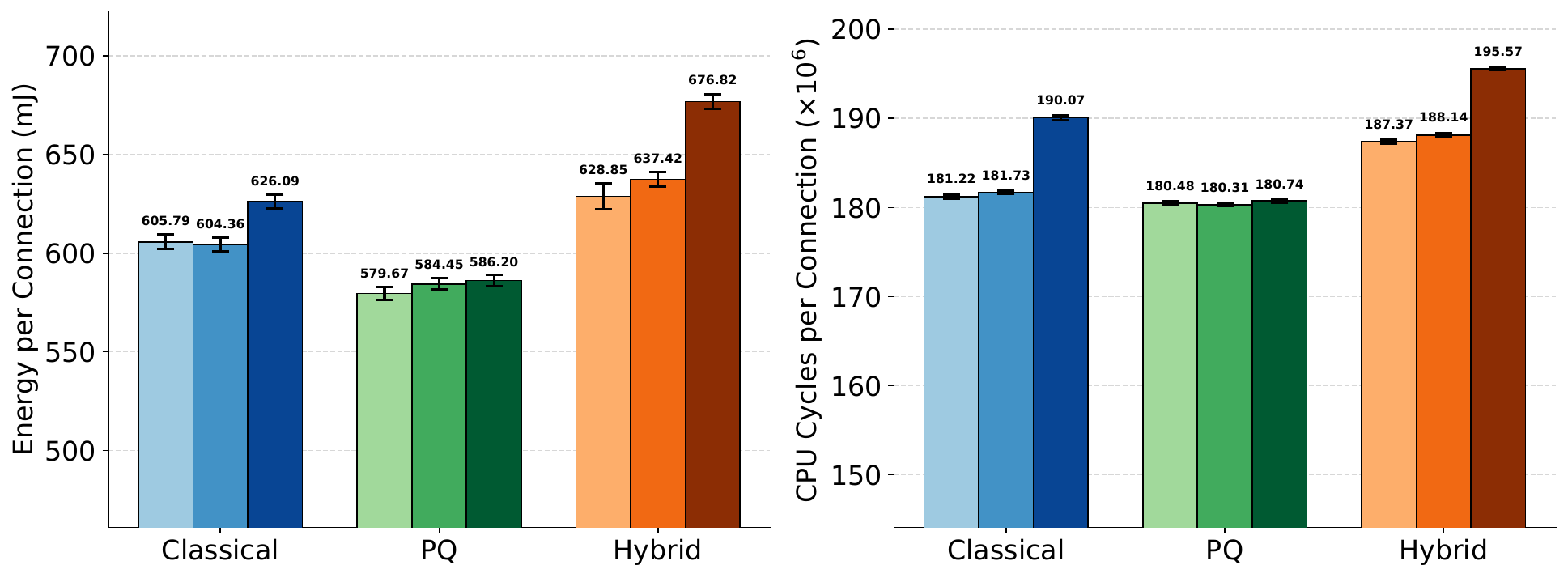}
\end{minipage}\hfill
\begin{minipage}[c]{0.55\textwidth}
    \centering
    \scriptsize
    \resizebox{\linewidth}{!}{%
    \begin{tabular}{|l|c|c|c|c|c|c|}
    \hline
    \multirow{2}{*}{Algorithm}
    & \multicolumn{2}{c|}{Delay=1ms}
    & \multicolumn{2}{c|}{Delay=10ms}
    & \multicolumn{2}{c|}{Delay=25ms} \\
    \cline{2-7}
    & Energy (mJ) & CPU ($\times10^6$) & Energy (mJ) & CPU ($\times10^6$) & Energy (mJ) & CPU ($\times10^6$) \\
    \hline
    X25519               & 432.21 & 68.27 & 1552.61 & 320.37 & 2549.84 & 736.86 \\
    SECP256r1            & 435.58 & 68.89 & 1498.26 & 321.22 & 2543.89 & 737.50 \\
    SECP384r1            & 489.66 & 78.19 & 1581.13 & 330.00 & 2558.88 & 746.39 \\
    ML-KEM-512           & 420.97 & 67.06 & 1403.68 & 319.71 & 2494.59 & 733.84 \\
    ML-KEM-768           & 422.59 & 67.20 & 1463.21 & 319.70 & 2512.53 & 733.72 \\
    ML-KEM-1024          & 424.61 & 67.49 & 1790.44 & 320.14 & 2518.23 & 733.73 \\
    X25519-ML-KEM-768    & 479.05 & 75.01 & 1464.11 & 327.02 & 2614.97 & 741.41 \\
    SECP256r1-ML-KEM-768 & 480.69 & 75.50 & 1470.62 & 327.25 & 2617.66 & 742.30 \\
    SECP384r1-ML-KEM-1024 & 527.45 & 81.73 & 1505.45 & 337.90 & 2637.30 & 754.11 \\
    \hline
    \end{tabular}%
    }
\end{minipage}

\vspace{12pt}

\begin{minipage}[c]{0.40\textwidth}
    \centering
    \textbf{(c)}\vspace{-12pt}\\
    \includegraphics[width=0.95\linewidth]{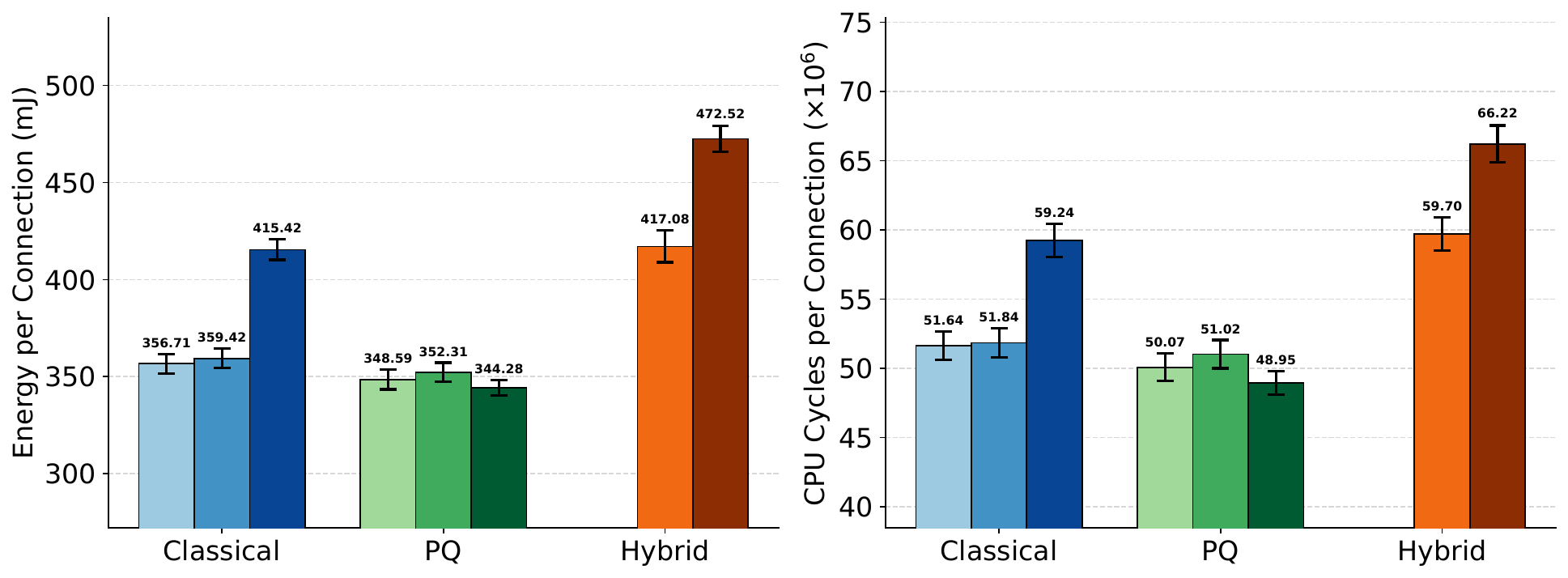}
\end{minipage}\hfill
\begin{minipage}[c]{0.55\textwidth}
    \centering
    \scriptsize
    \resizebox{\linewidth}{!}{%
    \begin{tabular}{|l|c|c|c|c|c|c|}
    \hline
    \multirow{2}{*}{Algorithm}
    & \multicolumn{2}{c|}{Loss=0.01\%}
    & \multicolumn{2}{c|}{Loss=1\%}
    & \multicolumn{2}{c|}{Loss=5\%} \\
    \cline{2-7}
    & Energy (mJ) & CPU ($\times10^6$) & Energy (mJ) & CPU ($\times10^6$) & Energy (mJ) & CPU ($\times10^6$) \\
    \hline
    X25519               & 300.99 & 42.60 & 588.92 & 137.24 & 1664.53 & 590.01 \\
    SECP256r1            & 306.10 & 43.16 & 588.83 & 137.50 & 1678.57 & 594.44 \\
    SECP384r1            & 361.94 & 51.05 & 647.40 & 147.09 & 1750.56 & 605.79 \\
    ML-KEM-512           & 292.82 & 41.19 & 559.37 & 128.62 & 1666.64 & 596.78 \\
    ML-KEM-768           & 294.02 & 41.57 & 588.33 & 139.40 & 1691.20 & 617.64 \\
    ML-KEM-1024          & 296.53 & 41.89 & 576.15 & 135.23 & 1740.50 & 701.03 \\
    X25519-ML-KEM-768    & 346.81 & 49.31 & 637.38 & 146.07 & 1851.87 & 711.74 \\
    SECP256r1-ML-KEM-768 & 351.38 & 49.92 & 612.23 & 135.47 & 1838.70 & 694.36 \\
    SECP384r1-ML-KEM-1024 & 408.59 & 57.30 & 718.25 & 161.23 & 1852.55 & 716.69 \\
    \hline
    \end{tabular}%
    }
\end{minipage}

\caption{Energy (mJ) and CPU cycles ($\times10^6$) per connection with ML-DSA-44 authentication. Layout as Fig.~\ref{fig:thr-all}: (a) MTU sweep, (b) delay sweep at 5\,ms, (c) loss sweep at 0.1\%.}
\label{fig:energy-all-mldsa44}
\end{figure*}

We evaluate nine KEM algorithms across three different categories. i) The classical baseline comprises X25519, SECP256r1, and SECP384r1. ii) The post-quantum group covers ML-KEM-512, ML-KEM-768, and ML-KEM-1024, corresponding to increasing security strengths, and iii) the hybrid group pairs classical ECDH with ML-KEM encapsulation: X25519+ML-KEM-768, SECP256r1+ML-KEM-768, and SECP384r1+ML-KEM-1024. All combinations use RSA-2048 and ML-DSA-44 as the authentication signature algorithm.


\noindent\textbf{Configuration Parameters}:
Each trial is parameterised by a (KEM, signature algorithm, network condition) tuple. For each combination, 50 independent runs are performed, each lasting 60\,s, yielding statistically stable estimates across all metrics. As described in Section~\ref{sec:Metrics}, the first 15 connections of every run serve as a warm-up phase and are excluded from all measurements.

Network parameters are varied one at a time. MTU is varied from 1200\,B to 9000\,B, one-way delay from 0 to 25\,ms, and packet-loss rate from 0 to 5\%. When one parameter is varied, the other two are held at their baseline values: MTU\,=\,1500\,B, delay\,=\,0\,ms, and loss\,=\,0\%.

\noindent\textbf{Results}:
\subsubsection{Throughput and Connection Time}
\indent \textit{a) Effect of MTU.:} Under ideal conditions (Fig.~\ref{fig:thr-all}a), pure ML-KEM variants marginally lead classical ECDH algorithms. ML-KEM-512 achieves 61.90\,conn/s, 2.5\% more than X25519 (60.38\,conn/s) while SECP384r1 lags 16\% behind ML-KEM-512, and SECP384r1+ML-KEM-1024 is 23.7\% lower in throughput (47.21 vs.\ 61.90\,conn/s) and 31\% slower in connection time (21.18 vs.\ 16.16\,ms) than ML-KEM-512, reflecting the accumulated cost of two heavyweight primitives. Sweeping MTU from 1200 to 9000~bytes produces no meaningful change: per-algorithm throughput varies by under 0.3~conn/s and connection time by under 0.2~ms across all MTU settings.

\indent \textit{b) Effect of Network Delay.:} With 1~ms one-way delay (Fig.~\ref{fig:thr-all}b), the best-to-worst throughput spread is 14\% (ML-KEM-512 at 40.36 vs.\ SECP384r1+ML-KEM-1024 at 34.71\,conn/s), and connection time differs by 16\% (24.78 vs.\ 28.81\,ms). As delay grows, RTT waiting dominates and cryptographic differences fade: by 25~ms, all algorithms converge to 4.15--4.25\,conn/s, a spread of under 2.4\%, rendering them essentially indistinguishable.

\indent \textit{c) Effect of Packet Loss.:} At 0.01\% loss (Fig.~\ref{fig:thr-all}c), the ranking mirrors ideal conditions: ML-KEM-768 leads at 60.82\,conn/s and SECP384r1+ML-KEM-1024 trails at 46.72\,conn/s, 23\% lower. Increasing loss to 1\% depresses all algorithms uniformly by 62--64\% (X25519: 59.78$\to$22.87; ML-KEM-512: 61.20$\to$22.21\,conn/s). At 5\% loss, hybrid connection times are up to 11\% longer than ML-KEM-512 (218.07 vs.\ 196.78\,ms for SECP384r1+ML-KEM-1024). The reason is TCP-level fragmentation: the 1{,}818-byte \texttt{ClientHello} of ML-KEM-1024 requires two TCP segments at MTU\,=\,1200, whereas ML-KEM-512 (1{,}050 bytes) fits within a single segment. Packet-trace measurements confirm this distinction quantitatively, yielding a mean retransmission count of 0.034 per handshake for ML-KEM-1024 at 1\% loss and 0.159 per handshake at 5\% loss, compared to zero at baseline. The 4.7-fold increase between these two loss rates, absent in single-segment algorithms, directly accounts for the disproportionate connection-time penalty observed in fragmented configurations.


\subsubsection{Energy and CPU Cycles}

\indent \textit{a) Effect of MTU.:} Under ideal conditions (Fig.~\ref{fig:energy-all}a), ML-KEM variants are the most efficient: ML-KEM-512 consumes 291.30\,mJ per connection, 2.6\% less than X25519 (299.12\,mJ), with 2.5\% fewer CPU cycles (40.37 vs.\ 41.39\,$\times10^6$). SECP384r1 is 24\% more energy-intensive than ML-KEM-512 (361.69 vs.\ 291.30\,mJ), and SECP384r1+ML-KEM-1024 incurs 41\% higher energy (411.70\,mJ) and 40\% more CPU cycles (56.58\,$\times10^6$) the direct arithmetic cost of running both primitives. Across MTU values, CPU cycle variation stays under 0.6\% for all algorithms.

\indent \textit{b) Effect of Network Delay.:} Idle socket waiting inflates the per-connection energy budget of all algorithms equally as delay rises, compressing inter-algorithm differences sharply (Fig.~\ref{fig:energy-all}b). At 1~ms delay, SECP384r1+ML-KEM-1024 consumes 24.5\% more energy (529.86 vs.\ 425.44\,mJ) than ML-KEM-512, versus 41\% under ideal conditions. By 25~ms, this gap collapses to under 5\% in energy (2641.79 vs.\ 2518.87\,mJ) and under 2.5\% in CPU cycles (754.06 vs.\ 735.47\,$\times10^6$), since idle wait time dominates over cryptographic computation entirely.

\indent \textit{c) Effect of Packet Loss.:} At 0.01\% loss (Fig.~\ref{fig:energy-all}c), SECP384r1+ML-KEM-1024 is 41\% more energy-intensive (404.89 vs.\ 287.07\,mJ) and 40\% higher in CPU cycles (57.53 vs.\ 41.05\,$\times10^6$) than ML-KEM-512. At 1\% loss, retransmissions shift the dominant cost away from cryptographic computation: SECP384r1+ML-KEM-1024 is 22\% more costly in energy (671.83 vs.\ 552.14\,mJ) and 15\% in CPU cycles (149.92 vs.\ 130.82\,$\times10^6$) than ML-KEM-768. At 5\% loss, X25519+ML-KEM-768 consumes 10\% more energy than X25519 alone (1820.77 vs.\ 1661.04\,mJ). The retransmission measurements establish the causal basis for this pattern: ML-KEM-1024 incurs 0.034 mean retransmissions per handshake at 1\% loss and 0.159 at 5\% loss. Each retransmission event requires TCP timeout expiry or fast retransmit recovery, during which the socket remains open and CPU cycles accumulate in kernel and application idle wait. The 4.7-fold increase in retransmission rate from 1\% to 5\% loss thus propagates into proportionally larger energy and CPU overhead for fragmented algorithms, an effect that is absent for single-segment algorithms such as ML-KEM-512 at the same loss rates.

\subsubsection{Effect of Authentication Algorithm}

The authentication algorithm governs the size of the server certificate chain and the \texttt{CertificateVerify} record. The ML-DSA-44 signature is 2{,}420 bytes, compared to approximately 256 bytes for RSA-2048, a difference that increases the total byte count of the server-to-client handshake flight and may add one or more TCP segments to that flight under constrained MTU settings.

Under zero-loss conditions, this size difference produces no statistically significant change in throughput or connection time. ML-KEM-512 achieves 61.90\,conn/s with RSA-2048 and 61.69\,conn/s with ML-DSA-44 at MTU\,=\,1200, a deviation of 0.3\%. The pattern holds across all nine KEM configurations and all three MTU values in Fig.~\ref{fig:thr-all-mldsa44}, confirming that the cryptographic computation cost of ML-DSA-44 is fully absorbed within the TLS processing pipeline and that the larger certificate size does not introduce a perceptible additional round-trip delay under clean-link conditions.

Under packet loss, the authentication algorithm becomes the dominant structural factor determining retransmission rate. Packet-trace measurements across all eighteen KEM--authentication combinations, conducted at MTU\,=\,1200 with packet-loss rates of 0\%, 1\%, and 5\%, establish the following pattern. At 0\% loss, all combinations produce zero retransmissions per handshake, confirming that fragmentation alone does not cause retransmissions in the absence of link impairment. At 1\% loss, RSA-2048 combinations yield a mean retransmission rate of 0.028--0.046 per handshake across KEM variants, while ML-DSA-44 combinations yield 0.060--0.075 per handshake, a factor of approximately 2.0--2.1$\times$ higher. At 5\% loss, this differential widens further: RSA-2048 combinations range from 0.136 to 0.234 retransmissions per handshake, while ML-DSA-44 combinations range from 0.307 to 0.388, a ratio of approximately 2.1--2.5$\times$. The consistent proportionality across loss levels confirms that the retransmission amplification is structural rather than stochastic: the larger ML-DSA-44 certificate places one additional TCP segment into the server-to-client flight, introducing an additional independent loss point at each loss rate.

Within each authentication class, the effect of KEM selection on retransmission rate is secondary. For RSA-2048, ML-KEM-512 (0.147) and ML-KEM-1024 (0.186) differ by a factor of 1.27$\times$  at 5\% loss, compared to the 2.1$\times$ gap between RSA-2048 and ML-DSA-44 at the same loss rate. The hybrid combinations incur the highest retransmission rates within each authentication class: SECP384r1+ML-KEM-1024 reaches 0.234 with RSA-2048 and 0.378 with ML-DSA-44 at 5\% loss, the highest values observed, because stacking a large classical key, a large KEM, and a large certificate maximises the total segment count of the handshake and thus the expected number of loss events per connection.

The results establish a two-tier fragmentation hierarchy. The primary split is between authentication algorithms, with ML-DSA-44 consistently incurring approximately twice the retransmission rate of RSA-2048 across all KEM configurations and loss conditions. The secondary split is between KEM variants within an authentication class, where larger KEMs modestly increase the retransmission rate relative to smaller ones. This ordering has a direct implication for deployment: in networks where packet loss exceeds 1\%, the choice of authentication algorithm has a greater impact on handshake reliability than the choice of KEM security level.

	\subsection{Deployment Implications}
\label{sec:deployment}

Algorithm selection for post-quantum TLS depends jointly on network conditions, handshake size, and the resulting fragmentation behavior~\cite{Bizri:PQCMigration:2026, Ogundipe:PQCThreat:2024}. The following scenarios use our measured performance data (\S~\ref{sec:experiments}) to provide concrete algorithmic recommendations for four deployment regimes.

\noindent\textbf{Datacenter and High-Speed LAN.}
In controlled data centre environments with sub-millisecond one-way delay and packet loss below 0.01\%, the TLS handshake overhead for all nine algorithm configurations is negligible relative to application-layer processing. Kampanakis and Childs-Klein~\cite{Kampanakis:DataHeavyTLS:2024}
report that the penalty of data-heavy PQ TLS on total transfer time falls below 5\% on stable, high-bandwidth links, consistent with our finding that ML-KEM-512 achieves 61.90\,conn/s at 291.30\,mJ per connection, marginally exceeding X25519 at 60.38\,conn/s and 299.12\,mJ. 
At this operating point, the size difference between standalone PQ and hybrid handshakes does not cross a fragmentation threshold, and the choice of authentication algorithm between RSA-2048 and ML-DSA-44 produces no measurable difference in throughput or energy.
\newline\noindent\textit{\textbf{Recommendation}: ML-KEM-512 with ML-DSA-44 achieves full NIST
compliance with no throughput or energy cost relative to the classical baseline.} Although all nine algorithm configurations perform equally at this operating point, a hybrid algorithm is not recommended for this regime: datacenter links are physically isolated private networks, where passive interception by an external attacker is not a realistic threat model. In this environment, the classical ECDH component of a hybrid handshake provides no additional security benefit, while its larger handshake size unnecessarily increases the risk of crossing a fragmentation threshold under reduced-MTU configurations. Standalone ML-KEM-512 therefore provides the optimal balance of post-quantum security and operational simplicity for intra-datacenter TLS.


\noindent\textbf{Enterprise Private Networks and 5G Campus Deployments.}
In enterprise 5G campus deployments, one-way delay falls between 4--20\,ms with low packet drop~\cite{Rischke:5GCampus:2021,Sathya:Enterprise5G:2024}. At 10\,ms delay, our measurements show throughput for all nine algorithms converges to 9.24--9.82\,conn/s, a spread of less than 6\%, as propagation delay dominates cryptographic computation differences. Connection-time separation measurable under ideal conditions becomes statistically indistinguishable at this delay level, and authentication algorithm choice contributes no differentiation below 0.1\% loss.\\
\textit{\textbf{Recommendation}: ML-KEM-768 provides NIST Level~3 security without throughput or energy disadvantage relative to ML-KEM-512 at these latencies, and is preferred where larger long-term security margins are required.} \\
\textbf{Mobile and LTE networks}: These networks impose RTTs of 15\,ms or more~\cite{Lauridsen:LTE5G:2017,Bennis:URLLC:2018}, with reliability degrading in mobile or rural scenarios~\cite{Zhang:5GEmpirical:2025}. In this regime, TCP-level handshake fragmentation is the primary performance differentiator: packet counts at NIST Level~3 rise from 4 at MTU\,=\,1500 to 7 at MTU\,=\,512~\cite{Abbasi:PQCHeterogeneous:2025}, with each additional fragment constituting an independent retransmission failure point~\cite{Claeys:TLS6TiSCH:2021}.
Our packet-trace measurements confirm the same causal structure in a
server-class TLS setting: at 1\% loss, ML-KEM-1024 incurs 0.034 mean retransmissions per handshake versus zero for single-segment algorithms,
rising to 0.159 at 5\% loss, a 4.7-fold amplification. Connection time for SECP384r1+ML-KEM-1024 reaches 218.07\,ms at 5\% loss versus 196.78\,ms for ML-KEM-512, an 11\% increase driven by this fragmentation exposure.
Measured mobile IIoT deployments using hybrid ML-KEM-768 report mean end-to-end latency around 108.8\,ms under representative conditions,
meeting sub-300\,ms industrial targets when persistent sessions are used to amortize handshake cost~\cite{Setyowati:MobileIIoT:2025}.
\newline\noindent\textit{\textbf{Recommendation}: Standalone ML-KEM-512 or ML-KEM-768 is preferred in loss-prone mobile environments; hybrid configurations impose a fragmentation-driven penalty that scales with packet loss rate without any compensating transport-layer benefit.}

\noindent\textbf{High-Security, Managed Low-Loss Links.}
Financial, governmental, and critical-infrastructure deployments requiring
NIST Level~5 assurance typically operate over managed, quality-assured
links where packet loss remains below 0.1\% and one-way delay is bounded.
At this operating point, SECP384r1+ML-KEM-1024 incurs a predictable and
bounded performance cost relative to smaller algorithms: throughput is
47.21\,conn/s, compared to 61.90\,conn/s for ML-KEM-512, a 23.7\%
reduction that is constant across all clean-link conditions tested and
does not amplify with network impairment in low-loss operation. Per-connection energy is 411.70\,mJ, 41.4\% above ML-KEM-512, consistent with findings that ML-KEM combined with ML-DSA offers the best latency-throughput profile in 5G user-equipment~\cite{Hoque:5GPQC:2025}.
\newline\noindent\textit{\textbf{Recommendation}: SECP384r1+ML-KEM-1024 with ML-DSA-44 is appropriate for this regime; the 23.7\% throughput reduction is deterministic and does not compound under the low-loss conditions expected in managed deployments.} 
\newline\noindent The 41.4\% higher per-connection energy is acknowledged as a real cost. However, organizations in this regime explicitly target NIST Security Category~5 equivalent in strength to AES-256 because their threat model requires the strongest available resistance to long-term quantum attacks on sensitive or classified data. For such deployments, the energy premium is a deliberate and accepted trade-off for the highest cryptographic assurance, and it remains a fixed, bounded, and plannable overhead that does not amplify under the stable low-loss links characteristic of managed infrastructure.


Across all four deployment scenarios, PQ-TLS performance cost separates into two independent components: a fixed cryptographic overhead that is network-condition-invariant, and a variable fragmentation penalty that scales with handshake size and link loss rate~\cite{Schoffel:EnergyIoT:2021}. Standalone ML-KEM configurations remain below the single-segment threshold across all standard MTU values and are deployment-ready without network qualification; hybrid configurations cross this threshold under reduced MTU, imposing a loss-sensitive penalty experimentally quantified in this work.
	
	\section{Conclusions}
	\label{sec:conclusion}

In this work, we evaluate the performance of classical, post-quantum, and hybrid key exchange algorithms in TLS 1.3. All measurements were conducted on identical hardware using the same methodology across 50 independent runs per configuration, ensuring that results are directly comparable across algorithm families. We focus on TLS handshake as it relies on asymmetric cryptography and is the layer most directly threatened by the advent of quantum computers.

Our results show that standalone post-quantum algorithms are fully competitive with classical ECDH. ML-KEM-512 matches and marginally exceeds X25519 in throughput and energy efficiency, and this holds across all tested MTU values with less than 0.3\,conn/s variation. We also found that scaling from ML-KEM-512 to higher security levels, ML-KEM-768 and ML-KEM-1024, incurs negligible overhead. These observations substantially weaken the commonly cited performance argument for delaying migration and suggest that performance concerns should no longer be a significant barrier to post-quantum migration. 

For hybrid schemes, the picture is more complex. The overhead of combining two key exchange primitives is measurable even under ideal conditions: SECP384r1+ML-KEM-1024 delivers roughly 24\% lower throughput and consumes 41\% more energy per connection than ML-KEM-512. In lossy networks this gap widens further, as larger hybrid payloads face higher retransmission probability. At 5\% packet loss, connection times for the heaviest hybrid run can be 11\% above the PQ only algorithm.

Our findings showed that network delay is a stronger determinant of performance than algorithm choice. Beyond 10ms of one-way latency, cryptographic differences shrink rapidly, and by 25ms all nine algorithms are within 2.4\% of each other in throughput. In conclusion, there is no significant performance barrier to adopting post-quantum TLS when using standalone ML-KEM algorithms. We expect our results to be useful for practitioners choosing algorithms for deployment. We hope the broader PQC community factors network-level performance into algorithm design, so that the security transition also remains an efficiency improvement for real deployments.

\section{Acknowledgements}
We thank funding support from Nokia Solutions and Networks India Private Limited, the Department of Telecommunications (DoT), and the Ministry of Electronics and Information TechnologY (MeitY)

	\bibliographystyle{IEEEtran}
	\bibliography{IEEEabrv, ref}

\end{document}